\documentclass[11pt,a4paper]{article}
\usepackage{amssymb}
\usepackage{amsmath}
\usepackage{latexsym}
\usepackage{graphicx} 
\usepackage{hyperref}
\usepackage{amsthm}
\usepackage{verbatim}
\usepackage{psfrag}
\usepackage{bbm}
\usepackage{dsfont}
\usepackage{mathrsfs}
\usepackage{tikz,pgfplots}
\usepackage{yfonts}
\usepackage{color}
\usepackage{enumerate}
\usepackage{subcaption}
\usepackage{caption}
\usepackage{pgfplots}
\usepackage{subcaption}

\newcommand*{\collauthor}[2]{{#1}$^{#2}$}

\newcommand*{\affiliation}[2]{$\mbox{}^{{#2}}${#1}}
\newcommand*{\colltitle}[1]{\textbf{#1}}

\newtheorem{pro}{Proposition}[section]
\newtheorem{theorem}[pro]{Theorem}
\newtheorem{remark}[pro]{Remark}
\newtheorem{lemma}{Lemma}[section]
\newenvironment{keywords}[1]{\vspace{1cm}\\{\bf \slshape{Keywords}}\quad\slshape{#1}}{}

\theoremstyle{definition}
\newtheorem{exmp}{Example}[section]

\begin{document}
\begin{center}
\begin{Large}
  \colltitle{Multistep schemes for solving backward stochastic differential equations on GPU}
\end{Large} 
\vspace*{1.5ex}

\begin{sc}
\begin{large}
\collauthor{Lorenc Kapllani}{},
\collauthor{Long Teng}{}
\end{large}
\end{sc}
\vspace{1.5ex}

\affiliation{Lehrstuhl f\"ur Angewandte Mathematik und Numerische Analysis,\\
Fakult\"at f\"ur Mathematik und Naturwissenschaften,\\
Bergische Universit\"at Wuppertal, Gau{\ss}str. 20, \\
42119 Wuppertal, Germany\linebreak }{} \\

\end{center}
\section*{Abstract}
The goal of this work is to parallelize the multistep scheme for solving the backward stochastic differential equations (BSDEs) in order to achieve both, a high accuracy and a reduction of the computation time as well. In the multistep scheme the computations at each grid point are independent and this fact motivates us to select massively parallel GPU computing using CUDA. In our investigations we identify performance bottlenecks and apply appropriate optimization techniques for reducing the computation time, using a uniform domain. Finally, some examples with financial applications are provided to demonstrate the achieved acceleration on GPUs.
\begin{keywords}
backward stochastic differential equations, multistep scheme, GPU computing, CUDA, option pricing
\end{keywords}

\section{Introduction}
In this work we parallelize the multistep scheme developed in~\cite{teng2018multi} to approximate numerically the solution of the following (decoupled) {\em forward backward stochastic differential equation (FBSDE)}:
\begin{equation}
   		  \begin{cases}
   		        \,\,dX_t =a\left(t,X_t\right)\,dt + b\left(t,X_t\right)\,dW_t, \quad X_0 = x_0,\\
   	   		-dy_t = f\left(t,X_t,y_t,z_t\right)\,dt -z_t\,dW_t,\\  
   		   		 \quad y_T = \xi = g\left(X_t\right),
   		   \end{cases}
  	\label{eq1}
\end{equation}
where $X_t, a \in \mathcal{R}^n $, $b$ is a $n\times d$ matrix, $W_t$ is a $d$-dimensional Brownian motion, $f\left(t,X_t,y_t,z_t\right):\left[0,T\right]\times\mathcal{R}^n\times\mathcal{R}^m\times\mathcal{R}^{m\times d} \to\mathcal{R}^m$ is the driver function and $\xi$ is the terminal condition. The terminal condition $y_T$ depends on the final value of a forward 
stochastic differential equation (SDE).
For $a = 0$ and $b = 1$, namely $X_t = W_t$, one obtains a 
{\em backward stochastic differential equation (BSDE)} of the form
\begin{equation}
   		  \begin{cases}
   		   		-dy_t= f\left(t,y_t,z_t\right)\,dt -z_t\,dW_t,\\  
   		   		 \quad y_T = \xi = g\left(W_T\right),
   		   \end{cases}
  	\label{eq2}
\end{equation}
where $y_t \in \mathcal{R}^m$ and $f\left(t,y_t,z_t\right):\left[0,T\right]\times\mathcal{R}^m\times\mathcal{R}^{m\times d} \to\mathcal{R}^m$. In the sequel of this work, we investigate the acceleration of numerical scheme developed in~\cite{teng2018multi} for solving \eqref{eq2}. Note that the developed schemes can be applied also for solving \eqref{eq1}, where the general Markovian diffusion $X_t$ can be approximated, e.g., by using the Euler-Scheme.

The existence and uniqueness of the solution of \eqref{eq2} are proven by Pardoux and Peng~\cite{Pardoux1990}. Peng~\cite{Peng1991} obtained a direct relation between FBSDEs and partial differential equations (PDEs). Based on this relationship, many numerical methods are proposed, e.g., probabilistic methods in~\cite{Bender2012,Bouchard2004,Gobet2005,Lemor2006,Zhao2006}, tree-based methods in~\cite{Crisan2012,Teng2018} etc. El Karoui et al.~\cite{El1997} showed that the solution of a linear BSDE is in fact the pricing and hedging strategy of an option derivative. This is the first claim of the application of BSDEs in finance.

In the field of financial mathematics, the approach with BSDEs has a couple of advantages compared to the standard approach with FSDEs. Firstly, many market models can be presented in terms of BSDEs (or FBSDEs), e.g., local volatility models~\cite{labart2011parallel}, stochastic volatility models~\cite{fahim2011probabilistic}, jump-diffusion models~\cite{eyraud2005backward}, defaultable options~\cite{ankirchner2010credit} etc. Secondly, BSDEs can also be used in incomplete markets~\cite{El1997} to solve the maximization problem of difference between the value of the super-replicating portfolio and the option value. Another advantage of using BSDEs is that one does not need to switch to the so-called risk-neutral measure for pricing financial options in complete markets. Therefore, BSDEs represent a more intuitive and more understandable way for pricing problems.

In general, the solution of BSDEs cannot be established in a closed form. Therefore, a numerical method is mandatory. There are two main classes of numerical methods for approximating the solution of BSDEs. The first class is proposed based on the relation between the BSDE and its related PDE. The other class contains approaches which are developed directly based on the BSDE. The $\theta$-discretization method has been most widely used, a second order convergence rate could be achieved with Crank Nicolson type. For a higher convergence rate, the authors in \cite{Zhao2010} proposed the multistep scheme in which the integrands are approximated by using Lagrange interpolating polynomials. For a better stability and the admission of more time levels, this multistep scheme has been generalised in \cite{teng2018multi} with spline instead of Lagrange interpolating polynomials. This kind of multistep schemes are computationally not efficient, since the values of the integrands at multiple time levels need to be known. Fortunately, computations in the multistep scheme are independent at each grid point. This fact motivates us to use massively parallel GPU computing to make these high-order accurate method more useful in practice.

Many acceleration strategies based on GPU computing have been developed for pricing problems, however, a very little of them are BSDE-based approach. In~\cite{Dai2010} a linear BSDE is solved on the GPU with the $\theta$-scheme method. They analyzed the effects of the thread number per block to increase the speedup. The parallel program with CUDA achieved high speedups and showed that the GPU architecture is well suited for solving the BSDEs in parallel. Peng et al.~\cite{Peng2011} developed acceleration strategies for option pricing with non-linear BSDEs using a binomial lattice based method. To increase the speedup, they reduce the global memory access frequency by avoiding the kernel invocation on each time step. Also, due to the load imbalance produced by the binomial grid, they provided load-balanced strategies and showed that the acceleration algorithms exhibit very high speedup over the sequential CPU implementation and therefore suitable for real-time application. Peng et al.~\cite{Peng2014} considered solving high dimensional BSDEs on GPUs with application in high dimensional American option pricing. A {\em Least Square Monte-Carlo (LSMC)} method based numerical algorithm is studied, and summarised in four phases. Multiple factors which affect the performance (task allocation, data store/access strategies and the thread synchronisation) are considered. Results showed much better performance than the CPU version. Gobet et al.~\cite{Gobet2016} designed a new algorithm for solving BSDEs based on LSMC. Due to stratification, the algorithm is very efficient especially for large scale simulations. They showed big speedups even in high dimensions.

Next we introduce some preliminaries needed to understand the multistep scheme. Let $\left(\Omega,\mathcal{F},\mathbb{P},\{\mathcal{F}_t\}_{0\le t \le T}\right)$ be a complete, filtered probability space. In this space a standard $d$-dimensional Brownian motion $W_t$ is defined, such that the filtration $\{\mathcal{F}_t\}_{0\le t\le T}$ is the natural filtration of $W_t.$ We define $|\cdot|$ as the standard Euclidean norm in the Euclidean space $\mathcal{R}^m$ or $\mathcal{R}^{m \times d}$ and $L^2 = L^2_{\mathcal{F}}\left(0,T; \mathcal{R}^d\right)$ the set of all $\mathcal{F}_t$-adapted and square integrable processes valued in $\mathcal{R}^d$. A pair of processes $\left(y_t,z_t\right):\left[0,T\right]\times\Omega\to\mathcal{R}^m \times \mathcal{R}^{m \times d}$ is the solution of BSDE \eqref{eq2} if it is $\mathcal{F}_t$-adapted, square integrable, and satisfies \eqref{eq2} in the sense of
\begin{equation}
	y_t = \xi + \int_t^T f\left(s, y_s, z_s\right)\,ds 
	          - \int_t^T z_s\,dW_s, \quad t\in \left[0,T\right),
  	\label{eq3}
\end{equation}
where $f\left(t,y_t,z_t\right):\left[0,T\right]\times \mathcal{R}^m\times\mathcal{R}^{m\times d} \to \mathcal{R}^m$ is $\mathcal{F}_t$-adapted and the third term on the right-hand side is an It\^o-type integral. This solution exist under regularity conditions~\cite{Pardoux1990}. Let us consider the following:
\begin{equation}
	y_t = u\left(t,W_t\right), \quad z_t=\nabla u\left(t,W_t\right) \quad \forall t \in \left[0,T\right),
\label{eq4}
\end{equation}
where $\nabla u$ denotes the derivative of $u\left(t,x\right)$ with respect to the spatial variable $x$ and $u\left(t,x\right)$ is the solution of the following (backward in time) parabolic PDE:
\begin{equation}
	\frac{\partial u}{\partial t} 
	+ \frac{1}{2} \sum_{i=1}^{d}\frac{\partial^2 u}{\partial x^2_i} + f\left(t,u,\nabla u\right) = 0,
\label{eq5}
\end{equation}
with the terminal condition $u\left(T,x\right)=\phi(x)$. 
Under regularity conditions, the PDE~\eqref{eq5} possess a unique solution $u\left(t,x\right)$. 
Therefore, for $\xi = \phi\left(W_T\right)$, the pair $\left(y_t, z_t\right)$ is the unique solution of BSDE~\eqref{eq3}.

Now we introduce some notation which is used in the Sections that follow. Let $\mathcal{F}_s^{t,x}$ for $t\le s\le T$ be a $\sigma$-field generated by the Brownian motion $\{x+W_r-W_t,t\le r\le s\}$ starting from the time-space point $(t,x)$. We define $E_s^{t,x}\bigl[X\bigr]$ as the conditional expectation of the random variable $X$ under the filtration $\mathcal{F}_s^{t,x}$, i.e.\ $E_s^{t,x}\bigl[X\bigr]=E\bigl[X\vert \mathcal{F}_s^{t,x}\bigr]$.

In the next Section, we introduce the multistep scheme. In Section~\ref{sec3}, we present our algorithmic framework. Section~\ref{sec4} is devoted to strategies of parallel GPU computing using CUDA. In Section~\ref{sec5}, we illustrate our findings with some examples including financial applications. Finally, Section~\ref{sec6} concludes this work.

\section{The multistep scheme}
\label{sec2}
In this Section we present the multistep scheme \cite{teng2018multi}.
\subsection{The stable semidiscrete scheme}
\label{subsec21}
Let $N$ be a positive integer and $\Delta t = T/N$ the step size that partitions uniformly the time interval $\left[0,T\right]$: $0 = t_0 < t_1 < \dots < t_{N-1} < t_N = T$,
where $t_n = t_0 + n\Delta t$, $n = 0,1,\dots,N$. 
Let $k$ and $K_y$ be two positive integers such that $1\le k \le K_y \le N$. The BSDE~\eqref{eq3} can be expressed as
\begin{equation}
	y_{t_n} = y_{t_{n+k}} 
     	+ \int_{t_{n}}^{t_{n+k}} f\left(s,y_s,z_s\right)\,ds
     	- \int_{t_{n}}^{t_{n+k}}z_s\,dW_s.
  	\label{eq6}
\end{equation}
In order to approximate $y_{t_n}$ based on the later information $\left[t_n,t_{n+k}\right]$, we need to obtain the adaptability. Therefore, we take the conditional expectation $E_{t_n}^x[\cdot]$ in \eqref{eq6} and obtain
\begin{equation}
	y_{t_n} = E_{t_n}^x\bigl[y_{t_{n+k}} \bigr] 
	      + \int _{t_n}^{t_{n+k}}E_{t_n}^x\bigl[f\left(s,y_s,z_s\right)\bigr]\,ds.
  	\label{eq7}
\end{equation}
In order to approximate the integral in \eqref{eq7}, Teng et al.~\cite{teng2018multi} used the cubic spline polynomial to approximate that integrand. Based on the support points $\left(t_{n+j}, E_{t_n}^x\bigl[f\left(t_{n+j},y_{t_{n+j}},z_{t_{n+j}}\right)\bigr]\right)$, $j=0,\cdots,K_y$, we have
\begin{equation}
    \int _{t_n}^{t_{n+k}}E_{t_n}^x\bigl[f\left(s,y_s,z_s\right)\bigr]\,ds = \int _{t_n}^{t_{n+k}}\tilde{S}_{K_y}^{t_n,x}\left(s\right)\,ds + R_y^n,
  	\label{eq8}
\end{equation}
where the cubic spline interpolant is given as
\begin{equation}
    \tilde{S}_{K_y}^{t_n,x}\left(s\right) = \sum_{j=0}^{K_y-1} \tilde{s}_{K_y}^{t_n,x,j}\left(s\right),
  	\label{eq9}
\end{equation}
where
$$\tilde{s}_{K_y}^{t_n,x,j}\left(s\right) = a_j^y+b_j^y\left(s-t_{n+j}\right)+c_j^y\left(s-t_{n+j}\right)^2+d_j^y\left(s-t_{n+j}\right)^3$$
with
$$ s \in \left[t_{n+j}, t_{n+j+1}\right], j = 0,\cdots,K_y-1.$$
Obviously, the residual reads
\begin{equation*}
    R_y^n = \int _{t_n}^{t_{n+k}} \bigl(E_{t_n}^x\bigl[f\left(s,y_s,z_s\right)\bigr] - \tilde{S}_{K_y}^{t_n,x}\left(s\right)\bigr)\,ds.
\end{equation*}
We calculate 
\begin{equation}
    \begin{split}
    \int _{t_n}^{t_{n+k}}\tilde{S}_{K_y}^{t_n,x}\left(s\right)\,ds &= \int _{t_n}^{t_{n+k}}\sum_{j=0}^{K_y-1} \tilde{s}_{K_y}^{t_n,x,j}\left(s\right)\,ds \\
    &= \sum_{j=0}^{K_y-1} \int _{t_n}^{t_{n+k}} \tilde{s}_{K_y}^{t_n,x,j}\left(s\right)\,ds \\
    &= \sum_{j=0}^{K_y-1} \int _{t_{n+j}}^{t_{n+j+1}} \tilde{s}_{K_y}^{t_n,x,j}\left(s\right)\,ds \\
    &=  \sum_{j=0}^{K_y-1} \left[a_j^y\Delta t+\frac{b_j^y\Delta t^2}{2}+\frac{c_j^y\Delta t^3}{3} + \frac{d_j^y\Delta t^4}{4}\right].
    \end{split}
    \label{eq10}
\end{equation}
and obtain the reference equation for $y$ as
\begin{equation}
	y_{t_n} = E_{t_n}^x\bigl[y_{t_{n+k}}\bigr] 
	 + \sum_{j=0}^{K_y-1} \left[a_j^y\Delta t+\frac{b_j^y\Delta t^2}{2}+\frac{c_j^y\Delta t^3}{3} + \frac{d_j^y\Delta t^4}{4}\right] + R_y^n.
  	\label{eq11}
\end{equation}

Let $\Delta W_s = W_s-W_{t_n}$ for $s \ge t_n$. Then $\Delta W_s$ is a standard Brownian motion with the zero mean and the standard deviation $\sqrt{s-t_n}$. Let $l$ and $K_z$ be two positive integers such that $1 \le l \le K_z \le N$. Using $l$ instead of $k$ in \eqref{eq6}, multiplying both sides by $\Delta W_{t_{n+l}}$, taking the conditional expectation $E_{t_n}^x[\cdot]$ and using the It\^o isometry we obtain
\begin{equation}
	0 = E_{t_n}^x\bigl[y_{t_{n+l}}\Delta W_{t_{n+l}} \bigr] 
	      + \int_{t_{n}}^{t_{n+l}} E_{t_n}^x\bigl[f\left(s,y_s,z_s\right) \Delta W_s\bigr]\,ds
	      - \int_{t_{n}}^{t_{n+l}} E_{t_n}^x\bigl[z_s \bigr]\,ds.
  	\label{eq12}
\end{equation}
Using again the cubic spline interpolation to approximate the two integrals in \eqref{eq12} and the relation 
$$ E_{t_n}^x\bigl[y_{t_{n+l}}\Delta W_{t_{n+l}} \bigr] = l\Delta tE_{t_n}^x\bigl[z_{t_{n+l}} \bigr],$$ we obtain the reference equation for $z$ process
\begin{equation}
    \begin{split}
	0 = l\Delta t E_{t_n}^x\bigl[z_{t_{n+l}} \bigr]
	 &+\sum_{j=0}^{K_z-1} \left[a_j^{z_1}\Delta t+\frac{b_j^{z_1}\Delta t^2}{2}+\frac{c_j^{z_1}\Delta t^3}{3} + \frac{d_j^{z_1}\Delta t^4}{4}\right] \\
	  &- \sum_{j=0}^{K_z-1} \left[a_j^{z_2}\Delta t+\frac{b_j^{z_2}\Delta t^2}{2}+\frac{c_j^{z_2}\Delta t^3}{3} + \frac{d_j^{z_2}\Delta t^4}{4}\right]+ R_{z_1}^n + R_{z_2}^n.
	 \end{split}
  	\label{eq13}
\end{equation}

The results above can be straightforwardly generalized to the $d$-dimensional case as
\begin{equation}
    \begin{split}
    	y_{t_n}^{\tilde{m}} &= E_{t_n}^x\bigl[y_{t_{n+k}}^{\tilde{m}}\bigr] 
	 + \sum_{j=0}^{K_y-1} \left[a_j^{y,\tilde{m}}\Delta t+\frac{b_j^{y,\tilde{m}}\Delta t^2}{2}+\frac{c_j^{y,\tilde{m}}\Delta t^3}{3} + \frac{d_j^{y,\tilde{m}}\Delta t^4}{4}\right] + R_y^{n,\tilde{m}},\\
	0 &= l\Delta t E_{t_n}^x\bigl[z_{t_{n+l}}^{\tilde{m},\tilde{d}} \bigr]
	 +\sum_{j=0}^{K_z-1} \left[a_j^{z_1,\tilde{m},\tilde{d}}\Delta t+\frac{b_j^{z_1,\tilde{m},\tilde{d}}\Delta t^2}{2}+\frac{c_j^{z_1,\tilde{m},\tilde{d}}\Delta t^3}{3} + \frac{d_j^{z_1,\tilde{m},\tilde{d}}\Delta t^4}{4}\right] \\
	  &- \sum_{j=0}^{K_z-1} \left[a_j^{z_2,\tilde{m},\tilde{d}}\Delta t+\frac{b_j^{z_2,\tilde{m},\tilde{d}}\Delta t^2}{2}+\frac{c_j^{z_2,\tilde{m},\tilde{d}}\Delta t^3}{3} + \frac{d_j^{z_2,\tilde{m},\tilde{d}}\Delta t^4}{4}\right]+ R_{z_1}^{n,\tilde{m},\tilde{d}} + R_{z_2}^{n,\tilde{m},\tilde{d}},
	 \end{split}
  	\label{eq14}
\end{equation}
where $\tilde{m}=1,2,\cdots,m$ and $\tilde{d}=1,2,\cdots,d$.

The unknown coefficients in~\eqref{eq14} are found using cubic spline conditions. For instance, for the $y$ process (in $1$-dimension), using support points $\left(t_{n+j}, E_{t_n}^x\bigl[f\left(t_{n+j},y_{t_{n+j}},z_{t_{n+j}}\right)\bigr]\right)$, $j=0,\cdots,K_y$, the conditions are
\begin{equation}
    \begin{cases}
    \quad \tilde{S}_{K_y}^{t_n,x}\left(t_{n+j}\right) =E_{t_n}^x\bigl[f\left(t_{n+j},y_{t_{n+j}},z_{t_{n+j}}\right)\bigr], \qquad j = 0,\cdots,K_y \\
    \, \, \, \,\tilde{s}_{K_y}^{t_n,x,j}\left(t_{n+j}\right) =\tilde{s}_{K_y}^{t_n,x,j+1}\left(t_{n+j}\right), \qquad j = 0,\cdots,K_y-2 \\
     \,\tilde{s}_{K_y}^{\prime \,\, t_n,x,j}\left(t_{n+j}\right) =\tilde{s}_{K_y}^{\prime \,\, t_n,x,j+1}\left(t_{n+j}\right), \qquad j = 0,\cdots,K_y-2 \\
    \tilde{s}_{K_y}^{\prime \prime \,\, t_n,x,j}\left(t_{n+j}\right) =\tilde{s}_{K_y}^{\prime \prime \,\, t_n,x,j+1}\left(t_{n+j}\right), \qquad j = 0,\cdots,K_y-2
    \end{cases}
    \label{eq15}
\end{equation}
For $K_y = 3$ and using not-a-knot cubic spline, the coefficients are calculated as follows. Consider the notation $$g_{t_{n+j}}=E_{t_n}^x\bigl[f\left(t_{n+j},y_{t_{n+j}},z_{t_{n+j}}\right)\bigr].$$ 
Then 
\begin{itemize}
    \item For $\tilde{s}_{K_y}^{t_n,x,0}\left(s\right), s \in \left[t_n, t_{n+1}\right]$
\end{itemize}
\begin{equation*}
    \begin{split}
        a_0 &= g_{t_{n}},\\
        b_0 &=-\left(11g_{t_{n}} - 18g_{t_{n+1}} + 9g_{t_{n+2}} - 2g_{t_{n+3}}\right)/6\Delta t, \\ 
        c_0 &=\left(2g_{t_{n}} - 5g_{t_{n+1}} + 4g_{t_{n+2}} - g_{t_{n+3}}\right)/2\Delta t^2, \\ 
        d_0 &=-\left(g_{t_{n}} - 3g_{t_{n+1}} + 3g_{t_{n+2}} - g_{t_{n+3}}\right)/6\Delta t^3,                      
    \end{split} 
\end{equation*}

\begin{itemize}
    \item For $\tilde{s}_{K_y}^{t_n,x,1}\left(s\right), s \in \left[t_{n+1}, t_{n+2}\right]$
\end{itemize}
\begin{equation*}
    \begin{split}
        a_1 &= g_{t_{n+1}},\\
        b_1 &=-\left(2g_{t_{n}} + 3g_{t_{n+1}} - 6g_{t_{n+2}} + g_{t_{n+3}}\right)/6\Delta t, \\ 
        c_1 &=\left(g_{t_{n}} - 2g_{t_{n+1}} + g_{t_{n+2}}\right)/2\Delta t^2, \\ 
        d_1 &=-\left(g_{t_{n}} - 3g_{t_{n+1}} + 3g_{t_{n+2}} - g_{t_{n+3}}\right)/6\Delta t^3,                  
    \end{split} 
\end{equation*}

\begin{itemize}
    \item For $\tilde{s}_{K_y}^{t_n,x,2}\left(s\right), s \in \left[t_{n+2}, t_{n+3}\right]$
\end{itemize}
\begin{equation*}
    \begin{split}
        a_2 &= g_{t_{n+2}},\\
        b_2 &=\left(g_{t_{n}} - 6g_{t_{n+1}} + 3g_{t_{n+2}} + 2g_{t_{n+3}}\right)/6\Delta t, \\ 
        c_2 &=\left(g_{t_{n}} - 2g_{t_{n+1}} + g_{t_{n+3}}\right)/2\Delta t^2, \\ 
        d_2 &=-\left(g_{t_{n}} - 3g_{t_{n+1}} + 3g_{t_{n+2}} - g_{t_{n+3}}\right)/6\Delta t^3.              
    \end{split} 
\end{equation*}
Reference equation for $y$ process can therefore be written as
\begin{equation*}
    \begin{split}
        y_{t_n} &= E_{t_n}^x\bigl[y_{t_{n+k}}\bigr] 
	 + \frac{3\Delta t}{8}g_{t_{n}} + \frac{9\Delta t}{8}g_{t_{n+1}}+\frac{9\Delta t}{8}g_{t_{n+2}} + \frac{3\Delta t}{8}g_{t_{n+3}} + R_y^n,\\
        &=E_{t_n}^x\bigl[y_{t_{n+k}}\bigr] + \Delta t K_y \sum_{j=0}^{K_y}\gamma _{K_y,j}^{K_y} E_{t_n}^x\bigl[f\left(t_{n+j},y_{t_{n+j}},z_{t_{n+j}}\right)\bigr] + R_y^n,                   
    \end{split} 
\end{equation*}
where 
$$\gamma _{K_y,0}^{K_y} = \gamma _{K_y,3}^{K_y}=\frac{1}{8}, \gamma _{K_y,1}^{K_y} = \gamma _{K_y,2}^{K_y}=\frac{3}{8}.$$
In a similar way, the corresponding coefficients can be found for other choices of $K_y.$ In \cite{teng2018multi}, the authors have shown that the semidiscrete scheme is stable when
\begin{equation*}
    k = 1,\cdots, K_y, ~\mbox{with}~ K_y = 1, 2, 3, \cdots , N, ~\mbox{and}~ l =1,  ~\mbox{with}~  K_z = 1, 2, 3, \cdots , N. 
\end{equation*}
This is to say that the algorithm allows for arbitrary multiple time levels $K_y$ and $K_z$. In Table~\ref{tab1} and~\ref{tab2}, we present the coefficients up to 6 time levels.

\begin{table}[h]
\centering
\caption{The coefficients $\bigl\{ \gamma_{K_y,j}^{K_y} \bigr\}_{j=0}^{K_y}$ until $K_y = 6$.}
\label{tab1}  
  \begin{tabular}{| c | c | c | c | c | c |c | c |}
    \hline
    $K_y$ & \multicolumn{7}{c|}{$\gamma_{K_y,j}^{K_y}$}\\ \hline
    & $j = 0$ & $j = 1$ & $j = 2$ & $j = 3$ & $j = 4$ & $j = 5$ & $j = 6$\\ \hline
    1 & $\frac{1}{2}$  & $\frac{1}{2}$ & & & & &\\ \hline
    2 & $\frac{1}{6}$  & $\frac{2}{3}$ & $\frac{1}{6}$ & & & &\\ \hline
    3 & $\frac{1}{8}$  & $\frac{3}{8}$ & $\frac{3}{8}$ & $\frac{1}{8}$ & & &\\ \hline
    4 & $\frac{1}{12}$  & $\frac{1}{3}$ & $\frac{1}{6}$ & $\frac{1}{3} $& $\frac{1}{12}$ & &\\ \hline
    5 & $\frac{41}{600}$  & $\frac{19}{75}$ & $\frac{107}{600}$ & $\frac{107}{600}$ & $\frac{19}{75}$ & $\frac{41}{600}$ &\\ \hline
    6 & $\frac{19}{336}$  & $\frac{3}{14}$ & $\frac{15}{112}$ & $\frac{4}{21}$ & $\frac{15}{112}$ & $\frac{3}{14}$ & $\frac{19}{336}$\\
    \hline
  \end{tabular}
\end{table}

\begin{table}[h]
\centering
\caption{The coefficients $\bigl\{ \gamma_{K_z,j}^{1} \bigr\}_{j=0}^{K_z}$ until $K_z = 6$.}
\label{tab2}  
  \begin{tabular}{| c | c | c | c | c | c |c | c |}
    \hline
    $K_z$ & \multicolumn{7}{c|}{$\gamma_{K_z,j}^{1}$}\\ \hline
    & $j = 0$ & $j = 1$ & $j = 2$ & $j = 3$ & $j = 4$ & $j = 5$ & $j = 6$\\ \hline
    1 & $\frac{1}{2}$  & $\frac{1}{2}$ & & & & &\\ \hline
    2 & $\frac{5}{12}$  & $\frac{2}{3}$ & $-\frac{1}{12}$ & & & &\\ \hline
    3 & $\frac{3}{8}$  & $\frac{19}{24}$ & $-\frac{5}{24}$ & $\frac{1}{24}$ & & &\\ \hline
    4 & $\frac{35}{96}$  & $\frac{5}{6}$ & $-\frac{13}{48}$ & $\frac{1}{12}$ & $-\frac{1}{96}$ & &\\ \hline
    5 & $\frac{131}{360}$  & $\frac{151}{180}$ & $-\frac{103}{360}$ & $\frac{37}{360}$ & $-\frac{1}{45}$ & $\frac{1}{360}$ &\\ \hline
    6 & $\frac{163}{448}$  & $\frac{47}{56}$ & $-\frac{129}{448}$ & $\frac{3}{28}$ & $-\frac{37}{1344}$ & $\frac{1}{168}$ & $-\frac{1}{1344}$\\
    \hline
  \end{tabular}
\end{table}

We denote $\left(y^n,z^n\right)$ as the approximation to $\left(y_{t_n},z_{t_n}\right)$,
 given random variables $\left(y^{N-i},z^{N-i}\right)$,  $i = 0,1,\dots,K-1$ 
with $K= \max\{K_y,K_z\}$, $\left(y^n,z^n\right)$ can be found for $n = N-K,\dots,0$ such that
\begin{equation}
	\begin{split}
		y^{n} &= E_{t_n}^x\bigl[y^{n+K_y} \bigr] + K_y\Delta t \sum_{j=0}^{K_y} \gamma_{K_y,j}^{K_y} \,E_{t_n}^x\bigl[f\left(t_{n+j},y^{n+j},z^{n+j}\right)\bigr] +R_y^n,\\
		0 & = E_{t_n}^x\bigl[z^{n+1} \bigr] + \sum_{j=1}^{K_z} \gamma_{K_z,j}^1 \, E_{t_n}^x\bigl[f\left(t_{n+j},y^{n+j},z^{n+j}\right)\Delta W_{t_{n+j}}^\top \bigr] - \sum_{j=0}^{K_z} \gamma_{K_z,j}^1 \,E_{t_n}^x\bigl[z^{n+j} \bigr]+\frac{R_z^n}{\Delta t},
	\end{split}
  	\label{eq16}
\end{equation}
where $y^{n} = \left(y^{n,1}, y^{n,2}, \cdots, y^{n,m}\right)^T$, $z^{n} = \left(z^{n,\tilde{m},\tilde{d}}\right)_{m\times d}$ and \\ $\Delta W_{t_{n+j}}^\top = \left(W_{t_{n+j}}^{1}, W_{t_{n+j}}^{2}, \cdots, W_{t_{n+j}}^{d}\right)^\top - \left(W_{t_{n}}^{1}, W_{t_{n}}^{2}, \cdots, W_{t_{n}}^{d}\right)^\top$. 
One can show that the local errors in \eqref{eq16} are given by
\begin{equation}
  |R_y^n|=\mathcal{O}\left(\Delta t^5\right), \qquad |R_z^n|=\mathcal{O}\left(\Delta t^5\right),
  \label{eq17}
\end{equation}
provided that $f$ and $g$ are smooth enough. In \eqref{eq15} we need to divide by $\Delta t$ to find the value of $z$ process. Therefore, in order to balance time truncation errors, one might set $K_z = K_y+1$. In the following, we only present the results of error analysis, for their proofs we refer to \cite{teng2018multi} and \cite{Zhao2010}. 
\begin{lemma}
The local estimates of the local truncation errors in \eqref{eq16} satisfy
\begin{equation*}
     |R_y^n|\le C\Delta t^{\min\left\{K_y+2,5\right\}}, \qquad |R_z^n|\le C\Delta t^{\min\left\{K_z+2,5\right\}},
\end{equation*}
where $C > 0$ is a constant depending on $T$, $f$, $g$ and the derivatives of $f$ and $g$.
\label{l1}
\end{lemma}
\begin{theorem}
Suppose that the initial values satisfy
\begin{equation*}
   		  \begin{cases}
   		       \max_{N-K_y+1\le n\le N}  E\bigl[|y_{t_n}-y^n|\bigr]=\mathcal{O}\left(\Delta t^{K_y+1}\right), for \,\,K_y = 1, 2, 3\\
   	   		    \max_{N-K_y+1\le n\le N}  E\bigl[|y_{t_n}-y^n|\bigr]=\mathcal{O}\left(\Delta t^{4}\right), for\,\, K_y >3 
   		   \end{cases}
\end{equation*}
for sufficiently small time step $\Delta t$ it can be shown that
\begin{equation}
      \sup_{0\le n\le N}  E\bigl[|y_{t_n}-y^n|\bigr] \le C\Delta t^{\min \left\{K_y+1, 4\right\}},
      \label{eq18}
\end{equation}
where $C > 0$ is a constant depending on $T$, $f$, $g$ and the derivatives of $f$ and $g$.
\label{th1}
\end{theorem}
\begin{theorem}
Suppose that the initial values satisfy
\begin{equation*}
   		  \begin{cases}
   		       \max_{N-K_z+1\le n\le N}  E\bigl[|z_{t_n}-z^n|\bigr]=\mathcal{O}\left(\Delta t^{K_z}\right), for \,\,K_z = 1, 2, 3\\
   	   		    \max_{N-K_z+1\le n\le N}  E\bigl[|z_{t_n}-z^n|\bigr]=\mathcal{O}\left(\Delta t^{3}\right), for\,\, K_z >3 
   		   \end{cases}
\end{equation*}
and the condition on the initial values in Theorem~\ref{th1} is fulfilled. For sufficiently small time step $\Delta t$ it can be shown that
\begin{equation}
      \sup_{0\le n\le N}  E\bigl[|z_{t_n}-z^n|\bigr] \le C\Delta t^{\min \left\{K_y+1, K_z, 3\right\}},
      \label{eq19}
\end{equation}
where $C > 0$ is a constant depending on $T$, $f$, $g$ and the derivatives of $f$ and $g$.
\label{th2}
\end{theorem}
\begin{remark}
    If $f$ does not depend on process $z$, the maximum order of convergence for $y$ process is $4$ and $3$ for $z$ process; If $f$ depends on process $z$, the maximum order of convergence for $y$ and $z$ processes is $3$
    \label{r1}
\end{remark}
\subsection{The fully discrete scheme}
\label{subsec22}
Let $\Delta x$ denote the step size in the partition of the uniform $d$-dimensional real axis, i.e.$$\mathcal{R}^{\tilde{d}} = \bigl\{ x_i^{\tilde{d}} \vert x_i^{\tilde{d}} \in \mathcal{R}, i\in \mathcal{Z}, x_i^{\tilde{d}} < x_{i+1}^{\tilde{d}}, \Delta x = x_{i+1}^{\tilde{d}} - x_{i}^{\tilde{d}}, \lim_{i\to+\infty} x_i^{\tilde{d}} = +\infty, \lim_{i\to-\infty} x_i^{\tilde{d}} = -\infty \bigr\},$$
where 
$$\mathcal{R}^{\tilde{d}} = \mathcal{R}^1 \times  \mathcal{R}^2 \times \cdots \times \mathcal{R}^d~\mbox{and}~\tilde{d} = 1, 2, \cdots , d.$$ Let $x_\bold{i} = \left(x^1_{i_
1}, x^2_{i_2}, \cdots, x^d_{i_d}\right)$ for $\bold{i} = \left(i_1, i_2, \cdots, i_d\right) \in \mathcal{Z}^d$.

We denote $\left(y^n_\bold{i},z^n_\bold{i}\right)$ as the approximation to $\left(y_{t_n,x_\bold{i}},z_{t_n,x_\bold{i}}\right)$,
given the random variables $(y^{N-l}_\bold{i},z^{N-l}_\bold{i})$, $l = 0,1,\dots,K-1$ with $K = \max\{K_y,K_z\}$, $(y^{n}_\bold{i},z^{n}_\bold{i})$ can be found for $n = N-K,\dots,0$ such that
\begin{equation}
    \begin{split}
        y^{n}_\bold{i} &= \hat{E}_{t_n}^{x_\bold{i}}\bigl[\hat{y}^{n+K_y}\bigr] 
		       + K_y\Delta t \sum_{j=1}^{K_y} b_{K_y,j}^{K_y}\, \hat{E}_{t_n}^{x_\bold{i}}\bigl[f(t_{n+j},\hat{y}^{n+j},\hat{z}^{n+j})\bigr]\\ 
		       &\qquad + K_y\Delta t b_{K_y,0}^{K_y}\,f(t_{n},y^{n}_{\bold{i}},z^{n}_{\bold{i}}),\\
		0 & = \hat{E}_{t_n}^{x_\bold{i}}\bigl[\hat{z}^{n+1} \bigr] 
		  + \sum_{j=1}^{K_z} b_{K_z,j}^1 \,\hat{E}_{t_n}^{x_\bold{i}}\bigl[f(t_{n+j},\hat{y}^{n+j},\hat{z}^{n+j}) \,\Delta W_{t_{n+j}}^\top \bigr]\\
		&\qquad - \sum_{j=1}^{K_z} b_{K_z,j}^1 \hat{E}_{t_n}^{x_\bold{i}}
		         \bigl[\hat{z}^{n+j} \bigr]-b_{K_z,0}^1 z^n_\bold{i}.
    \end{split}
    \label{eq20}
\end{equation}
where $\hat{E}_{t_n}^{x_i}\bigl[ \cdot \bigr]$ is used to denote the approximation of the conditional expectation. The functions in the conditional expectations involve the $d$-dimensional probability density function of the Brownian Motions, one can choose e.g., the Gauss-Hermite quadrature rule to achieve a high accuracy only with a few points. The conditional expectation can be approximated as
\begin{equation}
    \hat{E}_{t_n}^{x_\bold{i}}\bigl[\hat{y}^{n+k} \bigr] = \frac{1}{\pi ^\frac{d}{2}} \sum_{\Lambda=1}^{L}\omega_\Lambda \hat{y}^{n+k}\bigl(x_\bold{i}+\sqrt{2k\Delta t}\,a_\Lambda \bigr),    
\label{eq21}
\end{equation}
where $\hat{y}^{n+k}$ are interpolating values at the space points $\bigl(x_\bold{i}+\sqrt{2k\Delta t}\,a_\Lambda \bigr)$ based on $y^{n+k}$ values, $(\omega_\Lambda, a_\Lambda)$ for $\Lambda = \left(\lambda _1, \lambda _2,\cdots, \lambda _d\right)$ are the weights and roots of the Hermite polynomial of degree $L$ (\cite{abramowitz1972handbook}), $\omega_\Lambda = \prod_{\tilde{d}=1}^{d} \omega _{\lambda_{\tilde{d}}} $, $a_\Lambda =\left(  a_{\lambda _1}, a_{\lambda _2},\cdots, a_{\lambda _d}\right)$ and $\sum_{\Lambda=1}^{L} = \sum_{\lambda _1=1,\cdots, \lambda _d=1}^{L,\cdots , L}$. In the same way, one can express the other conditional expectations in \eqref{eq20}. 
	
\section{The algorithmic framework}\label{sec3}
In this Section we present the algorithmic framework of the proposed numerical method.
\subsection{The Algorithm}
\label{subsec31}
According to~\eqref{eq20}, we will consider the following three steps.

\begin{enumerate}
\item \textbf{Construct the time-space discrete domain.} \\
We divide the time period $[0,T]$ into $N$ time steps using 
$\Delta t = T/N$ and get $N+1$ time layers and the space domain~$\mathcal{R}^d$ as explained in Subsection~\ref{subsec22} using step size $\Delta x$. We will use the truncated domains $\left[-8, 8\right] \text{or} \left[-16, 16\right]$. Furthermore, in order to balance the errors in time and space directions, we adjust $\Delta x$ and $\Delta t$ such that they satisfy the equality 
$(\Delta x)^r=(\Delta t)^{q+1}$, 
where $q = \min\left\{K_y+1,K_z\right\}$ and $r$ denotes the global error from the interpolation method used to generate the non-grid points when calculating the conditional expectations.

\item \textbf{Calculate $K$ initial solutions with $K = \max\{K_y,K_z\}$.}\\
The terminal value is given, one needs to compute the other $K-1$ initial values. This can be done by running a $1$-step scheme for $[t_{N-K+1},t_{N-1}]$ with a sufficiently small $\Delta t$ such that the $K-1$ produced initial values will have neglectable error.

\item \textbf{Calculate the numerical solution $(y_0^0,z_0^0)$ backward using equation~\eqref{eq20}.}\\
Note that the calculation for the $y$ process is done implicitly with Picard iteration. 
\end{enumerate}

\subsection{Preliminary considerations}
\label{subsec32}
For our numerical experiments we give the following remarks:
\begin{itemize}
\item When generating the non-grid points for the calculation of conditional expectations, some of them will be outside of the truncated domain. For these points, we take the values on the boundaries.
\item Due to uniformity of the grid, one does not need to consider $2K$ ($K$ for $y$ and $K$ for $z$) interpolations for each new calculation, but only $2$. Suppose that we are at time layer $t_{n-K}$. 
To calculate $y$ and $z$ values on this time layer, one needs the calculation of conditional expectations for $K$ time layers. The cubic spline interpolation is used to find the non-grid values for $1$-dimensional cases and bicubic interpolation for $2$-dimensional cases. For instance, the coefficients for the $y$ process are $A_y\in\mathcal{R}^{K\times \left(4^d \times M^d\right)}$, all the coefficients are stored. When we are at time layer $t_{n-K-1}$, only the spline interpolation corresponding to the previous calculated values is considered. Then, the columns of matrix $A_y$ are shifted $+1$ to the right in order to delete the last column and enter the current calculated coefficients in the first column. 
The new $A_y$ is used for the current step. 
The same procedure is followed until $t_0$. 
This reduces the amount of work for the algorithm.
\item There is a very important benefit from the uniformity of the grid. When we need to find the position of the non-grid points, a naive search algorithm is to loop over the grid points. In the worst case, an $\mathcal{O}\left(M^d\right)$ work is needed. Fortunately, this can be done in $\mathcal{O}\left(d\right)$, i.e., without for-loop. Recall that each new point is generated as 
$X_{\lambda_{\tilde{d}}} = x_{i_{\tilde{d}}} + \sqrt{2\Delta t k}\, a_{\lambda_{\tilde{d}}}$. This means that taking $\text{int}\left(\frac{X_{\lambda_{\tilde{d}}}-x_{min}}{\Delta x}\right)$ for $x_{i_{\tilde{d}}} \in \left[x_{min}, x_{max}\right]$ and $M-\text{int}\left(\frac{X_{\lambda_{\tilde{d}}}-x_{min}}{\Delta x}\right)$ for $x_{i_{\tilde{d}}} \in \left[x_{max}, x_{min}\right]$ gives the left boundary of the grid interval that $X_{\lambda_{\tilde{d}}}$ belongs to. This reduces substantially the total computation time, as it will be demonstrated in the numerical experiments.
\end{itemize}

\subsection{The Parallel implementation}
\label{subsec33}
In this Section we present the naive parallelization of the multistep scheme. Nevertheless, we have kept into attention the optimal CUDA execution model, i.e., creating arrays such that the access will be aligned and coalesced, reducing the redundant access to global memory, using registers when needed etc.

The first and second steps of the algorithm are implemented in the host. The third step is fully implemented in the device. Recall from \eqref{eq20} that the following steps are needed to calculate the approximated values on each time layer backward:
\begin{itemize}
\item \textbf{Generation of non-grid points $X_\Lambda = x_{\bold{i}} + \sqrt{2\Delta t k}\, a_\Lambda$.} \\
In the uniform domain, the non-grid points need to be generated only once. 
To do this, a kernel is created where each thread generates $L^d$ points for each space direction. 

\item \textbf{Calculation of the values $\hat{y}$ and $\hat{z}$ at the non-grid points.} \\
This is the most time consuming part of the algorithm. For the $1$-dimensional cases, we have considered the cubic spline interpolation. Since~\eqref{eq20} involves the solution of two linear systems, the $BiCGSTAB$ iterative method is used since the matrix is tridiagonal. To apply the method, we consider the $cuBLAS$ and $cuSPARSE$ libraries. For the inner product, second norm and addition of vectors, we use the $cuBLAS$ library. For the matrix vector multiplication, we use the $cuSPARSE$ library with the compressed sparse row format, due to the structure of the system matrix. Moreover, we created a kernel to calculate the spline coefficients based on the solved systems. Finally, a kernel to apply the last point in Subsection~\ref{subsec32} is created to find the values at non-grid points. Note that each thread is assigned to find $m+m\times d$ values ($m$ for $y$ and $m\times d$ for $z$). For the $2$-dimensional examples, we have considered the bicubic interpolation. We need to calculate $16$ coefficients for each point. Based on the bicubic interpolation idea, we need the first and mixed derivatives. These are approximated using finite difference schemes of the fourth order of accuracy (central, forward and backward). Therefore, a kernel is created where each thread calculates these values. Moreover, to find the $16$ coefficients,  a matrix vector multiplication needs to be applied for each point. Therefore, each thread performs a matrix-vector multiplication using another kernel. Finally, a kernel to apply the last point in Subsection~\ref{subsec32} is created to find the values at non-grid points, where each thread calculates  $m+m\times d$ values. 

\item \textbf{Calculation of the conditional expectations.}\\
For the first conditional expectations in the right hand side of \eqref{eq20}, we creat one kernel, where each thread calculates one value by using \eqref{eq21}. 
Furthermore, we merged the calculation of three conditional expectation in one kernel, namely 
\begin{equation*}
    \hat{E}_{t_n}^{x_{\bold{i}}}\bigl[\hat{z}^{n+j} \bigr],\quad
    \hat{E}_{t_n}^{x_{\bold{i}}}\bigl[f(t_{n+j},\hat{y}^{n+j},\hat{z}^{n+j})\bigr], \quad \hat{E}_{t_n}^{x_{\bold{i}}}\bigl[f(t_{n+j},\hat{y}^{n+j},\hat{z}^{n+j}) \,\Delta W_{t_{n+j}}\bigr],
\end{equation*} 
for $j=1,2,\dots,K$. This reduces the accessing of data multiple times from the global memory. 
Note that one thread  calculates $2\times m\times d + m$ values.

\item \textbf{Calculation of the $z$ values.}\\
The second equation in \eqref{eq20} is used and each thread calculates $m\times d$ values.

\item \textbf{Calculation of the $y$ values.}\\
The first equation in \eqref{eq20} is used and each thread calculates $m$ values, using the Picard iterative process.
\end{itemize}

\section{GPU computing and CUDA}
\label{sec4}
In this Section, we discuss about GPU computing using CUDA. We start with CUDA programming and execution model and present an iterative process to optimize CUDA application. 

\subsection{CUDA programming and execution model}
\label{subsec41}
CUDA (Compute Unified Device Architecture) provides a framework for developing parallel general purpose applications on a GPU.
At its core, there are three key abstractions: a hierarchy of thread groups, a hierarchy of memory and multiple thread level communication. These abstractions provide granular and coarse parallelism. Therefore, the application domain can be divided into sub-domains based on the data independence. Threads are organised into blocks of threads (threads within the block can communicate), grid of blocks and are executed in the SIMD fashion (a group of 32 threads called warps). 
Since the execution is based on warps (and scheduled from the warp schedulers; each GPU has a number of warp schedulers), the dimension of thread blocks gives different performances.

In a GPU, the largest and slowest memory is the global memory, which allows us to transfer data between the host (CPU) and the device (GPU) and is accessible for all threads. The shared memory is exclusive to thread blocks. The access is significantly faster than that of the global memory, and using this memory is profitable for optimisation tuning.  
In order to have an optimal application, each kernel created should be checked for the possible performance limiters. After the performance inhibitor is found, different techniques are considered to overcome the problem. Note that there can be a trade-off between different techniques. In the next Subsection, we present an iterative process to optimize the performance of the application in the GPU.

\subsection{Iterative optimization process}
\label{subsec42}
The first version of a CUDA program is mostly not the optimal one. Therefore, we should access and identify the bottlenecks. There are three main limiting factors, memory bound, compute bound and latency bound. Therefore, we need to focus on efficient use of GPU memory bandwidth, compute resources and hiding of instruction and memory latency. To identify these factors, one can use CUDA profiling tools (NVIDIA Command-line Profiler or $nvprof$ and Visual NVIDIA profiler or $nvvp$). Profile-driven optimization is an iterative process to optimize the program based on profile information. We have used the following iterative approach:
\begin{enumerate}
    \item Apply profiler to the application to gather information
    \item Identify application hotspots 
    \item Determine performance inhibitors
    \item Optimize the code
    \item Repeat the previous steps until desired performance is achieved
\end{enumerate}

The $nvprof$ profiling tool enables the collection of a timeline of CUDA-related activities on both CPU and GPU, including kernel execution, memory transfers, memory set, CUDA API calls and events or metrics for CUDA kernels. Profiling options are provided through command-line. It is used for the first and second step of profile optimization process. The $nvvp$ is a graphical tool with two main features, a timeline to display CPU and GPU activity and automatic performance analysis to help in identifying optimization opportunities. It provides a guided analysis, and guides one step-by-step through analysis in the entire application. In this mode, it helps on understanding of the likely performance limiters and optimization opportunities, including CUDA application analysis, performance-critical kernels, compute, bandwidth or latency bound and compute resources. It is used for the third profile optimization process.

After that the performance inhibitor is found, we consider different techniques to overcome the problem (step $4$). The techniques usually are related with exposing sufficient parallelizm, optimizing memory access and optimizing instruction execution. There are two ways to increase parallelism: keeping more concurrent warps active within an Streaming Multiprocessor (SM) and assigning more independent work to each thread or warp. To keep more concurrent warps, we change the grid configuration (e.g. by decreasing the block dimension, one can have more blocks per SM etc.). To assign more independent work, we use unrolling techniques (an operation is split into multiple operations). Note that an increase of parallelism can be limited by compute resources such as shared memory and registers. A $100\%$ occupancy can't be reached in such case. Therefore, a trade of must be found. The goal of memory access optimization is to maximize memory bandwidth utilization, with the focus on memory access patterns (maximize the use of bytes that travel on the bus) and sufficient concurrent memory accesses (hide memory latency). The best access pattern to global memory is aligned and coalesced access. There are several ways to optimize instruction execution, including hiding latency by keeping sufficient active warps or assigning more independent work to a thread and avoiding divergent execution paths within a warp. For the first two, we can use the same techniques as in exposing sufficient parallelism. For diverges branches, CUDA has compiler optimization features that replaces branch instructions (which cause actual control flow to diverge) with predicated instructions. However, for a long code path, the warp divergence will happen. We should use the branches as less as possible and recall the SIMT execution type used by GPU.

In the next Section, we present some numerical examples to show the high effectiveness and accuracy of the numerical scheme using GPU computing.

\section{Numerical results}
\label{sec5}
The first example is a linear BSDE with the driver function $f$ that does not depending on the process $z$. The second one is a non-linear example. Furthermore, we consider the Black Scholes BSDE as an application of BSDEs in finance. Finally, we test our algorithm with a $2$-dimensional example and the zero strike European spread option. We implement the parallel algorithm using CUDA C programming. The parallel computing times are compared with the serial ones on a CPU. Furthermore, the speedups are calculated. The CPU is Intel(R) Core(TM) i$5$-$4670$ $3.40$Ghz with $4$ cores. The GPU is a NVIDIA GeForce $1070$ Ti with a total $8$GB GDDR$5$ memory.
\begin{exmp}
Consider the linear BSDE\footnote{Taken from \cite{Zhao2010}.}
\begin{equation}
   		  \begin{cases}
   		   		-dy_t = \bigl(-y_t^3 + \frac{5}{2} y_t^2 -\frac{3}{2} y_t \bigr)\,dt -z_t \,dW_t,\\  
   		   		 \quad y_T = \frac{\exp(W_T+T)}{\exp(W_T+T)+1}.
   		   \end{cases}
  	\label{eq22}
\end{equation}
\label{ex1}
\end{exmp}
The analytic solution is 
\begin{equation}
   		  \begin{cases}
   		   		y_t = \frac{\exp(W_t+t)}{\exp(W_t+t)+1},\\  
   		   		 z_t = \frac{\exp(W_t+t)}{\bigl(\exp(W_t+t)+1\bigr)^2}.
   		   \end{cases}
  	\label{eq23}
\end{equation}
The exact solution with $T=1$ is $\left(y_0,z_0\right)=\left(\frac{1}{2},\frac{1}{4}\right)$. 
In Table~\ref{tab3}, we show the importance of working in a uniform domain. Note that the computation time is in seconds.

\begin{table}[h]
\centering
\caption{Preliminary results for $N = 256, K_y = K_z = 3$.}
\label{tab3}      
\begin{tabular}{| c | c | c | c |}
\hline 
M & $t_{CPU}^{non-optimal}$ & $t_{CPU}^{optimal}$ & speedup  \\ \hline
8192 & 2041.89 & 11.02 & 185.31 \\ 
\hline
\end{tabular}
\end{table}

\begin{figure}[h!]
	\centering
	\begin{subfigure}[h!]{0.32\linewidth}
		\includegraphics[width=\linewidth]{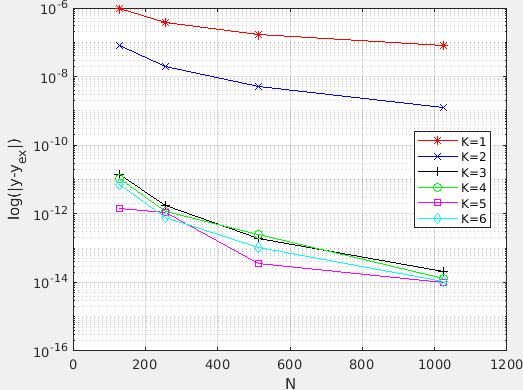}
		\caption{Plot of $y$ error as a function of time layers N for $K = 1,\cdots,6$.}
		\label{fig1a}
	\end{subfigure}
	\begin{subfigure}[h!]{0.32\textwidth}
		\includegraphics[width=\textwidth]{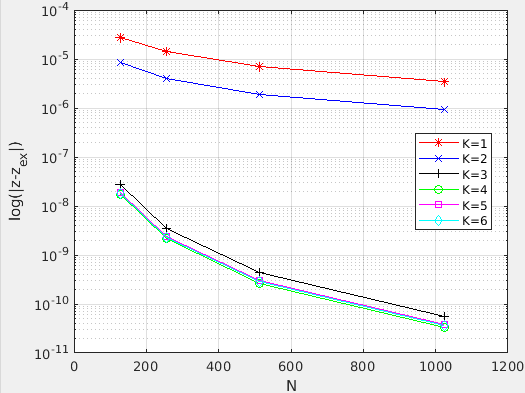}
		\caption{Plot of $z$ error as a function of time layers N for $K = 1,\cdots,6$.}
		\label{fig1b}
	\end{subfigure}
	\begin{subfigure}[h!]{0.32\textwidth}
		\includegraphics[width = \textwidth]{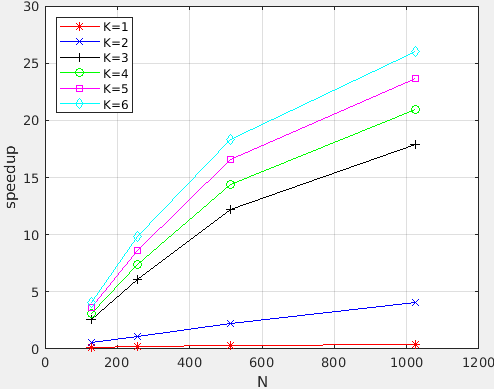}
		\caption{Plot of speedup as a function of time layers N for $K= 1,\cdots,6$.}
		\label{fig1c}
	\end{subfigure}
	\caption{Plots of naive results for Example~\ref{ex1}.}
	\label{fig1}
\end{figure}

\begin{figure}[h!]
	\centering
	\begin{subfigure}[h!]{0.32\linewidth}
		\includegraphics[width=\linewidth]{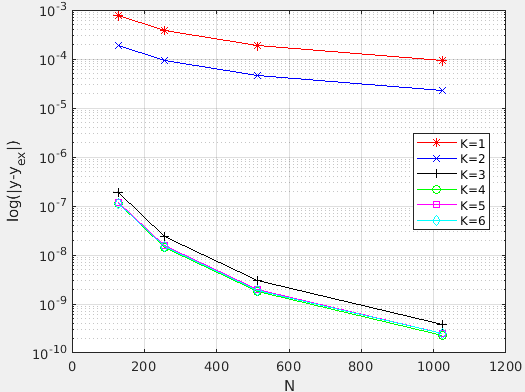}
		\caption{Plot of $y$ error as a function of time layers N for $K = 1,\cdots,6$.}
		\label{fig2a}
	\end{subfigure}
	\begin{subfigure}[h!]{0.32\textwidth}
		\includegraphics[width=\textwidth]{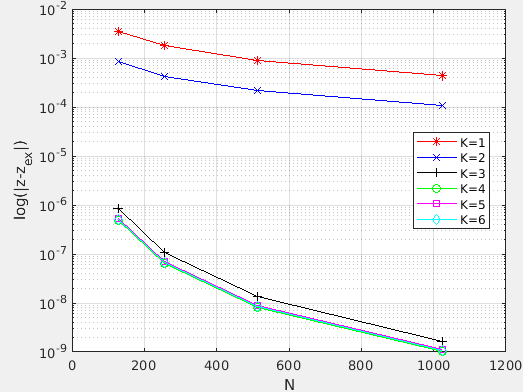}
		\caption{Plot of $z$ error as a function of time layers N for $K = 1,\cdots,6$.}
		\label{fig2b}
	\end{subfigure}
	\begin{subfigure}[h!]{0.32\textwidth}
		\includegraphics[width = \textwidth]{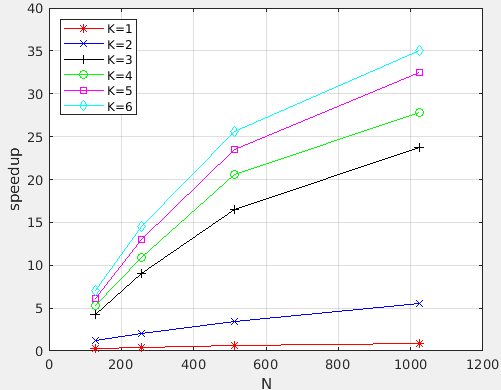}
		\caption{Plot of speedup as a function of time layers N for $K= 1,\cdots,6$.}
		\label{fig2c}
	\end{subfigure}
	\caption{Plots of naive results for Example~\ref{ex2}.}
	\label{fig2}
\end{figure}

In Table~\ref{tab4}, we present the naive results using $256$ threads per block with $K = K_y = K_z$, $t_0=0$, $T = 1$, $x \in [-16, 16]$, $L = 32$ and $p\footnote{Number of Picard iterations.} = 30$. For an easier understanding, the same results are plotted and presented in Figure~\ref{fig1}.
\begin{table}[h!]
\centering
\caption{Naive results for Example~\ref{ex1}.}
\label{tab4}  
  \begin{tabular}{| c | c | c | c | c | c | c |c |}
    \hline
    K & N & M & $|y_{0,0}-y_0^0|$ & $|z_{0,0}-z_0^0|$ & $t_{CPU}$ & $t_{GPU}$ & speedup\\ \hline
    1  & 128 & 364 & 9.36E-07 & 2.78E-05 & 0.14 & 0.91 & 0.15 \\ \hline
    1 & 256 & 512 & 3.89E-07 & 1.40E-05 & 0.37 & 1.73 & 0.21\\ \hline
    1 & 512 & 726 & 1.74E-07 & 7.04E-06 & 1.06 & 3.57 & 0.30\\ \hline
    1 & 1024 & 1024	& 8.22E-08 & 3.53E-06 & 2.91 & 6.96 & 0.42\\ \hline
    2 & 128 & 1218 & 8.01E-08 & 8.61E-06 & 0.64 & 1.05 & 0.61 \\ \hline 
    2 & 256 & 2048 & 2.03E-08 & 4.00E-06 & 2.06 & 1.88 & 1.10 \\ \hline
    2 & 512 & 3446 & 5.02E-09 & 1.92E-06 & 7.18 & 3.21 & 2.24 \\ \hline 
    2 & 1024 & 5794 & 1.25E-09 & 9.41E-07 & 23.93 & 5.83 & 4.10 \\ \hline
    3 & 128 & 4096 & 1.44E-11 & 2.77E-08 & 2.71 & 1.04 & 2.61 \\ \hline 
    3 & 256 & 8192 & 1.70E-12 & 3.50E-09 & 11.02 & 1.82 & 6.06 \\ \hline
    3 & 512 & 16384	& 1.87E-13 & 4.41E-10 & 44.86 & 3.68 & 12.19 \\ \hline 
    3 & 1024 & 32768 & 2.05E-14 & 5.53E-11 & 180.30 & 10.08 & 17.89 \\ \hline
    4 & 128 & 4096 & 1.06E-11 & 1.69E-08 & 3.28 & 1.05 & 3.13 \\ \hline
    4 & 256 & 8192 & 1.20E-12 & 2.13E-09 & 13.57 & 1.84 & 7.36 \\ \hline
    4 & 512 & 16384 & 2.57E-13 & 2.68E-10 & 55.16 & 3.84 & 14.35 \\ \hline
    4 & 1024 & 32768 & 1.29E-14 & 3.34E-11 & 223.28 & 10.68 & 20.91 \\ \hline
    5 & 128 & 4096 & 1.46E-12 & 1.90E-08 & 3.86 & 1.06 & 3.63 \\ \hline 
    5 & 256 & 8192 & 1.12E-12 & 2.40E-09 & 16.23 & 1.88 & 8.65 \\ \hline
    5 & 512 & 16384 & 3.46E-14 & 3.02E-10 & 65.80 & 3.97 & 16.57 \\ \hline
    5 & 1024 & 32768 & 9.77E-15 & 3.78E-11 & 267.79 & 11.33 & 23.64 \\ \hline
    6 & 128 & 4096 & 6.94E-12 & 1.84E-08 & 4.53 & 1.10 & 4.11 \\ \hline
    6 & 256 & 8192 & 7.71E-13 & 2.32E-09 & 18.97 & 1.93 & 9.84 \\ \hline
    6 & 512 & 16384 & 1.07E-13 & 2.92E-10 & 77.64 & 4.23 & 18.35 \\ \hline
    6 & 1024 & 32768 & 1.03E-14 & 3.65E-11 & 311.87 & 11.97 & 26.06 \\
    \hline
  \end{tabular}
\end{table}
It can be easily observed the increase of accuracy when considering a higher-step scheme. Since we have more time layers to consider, more work can be assigned to the GPU and therefore the speedup of the application is increased. That is why the highest speedup ($26\times$) is for a $6$-step scheme. Also the highest accuracy is of $\mathcal{O}\left(10^{-15}\right)$ for the $y$ process, since it has $4$-th order of convergence.
\begin{exmp}
Consider the non-linear BSDE\footnote{Taken from \cite{Zhao2010}.}
\begin{equation}
   		  \begin{cases}
   		   		-dy_t= \frac{1}{2} \left( \exp\left( t^2 \right) -4ty_t-3 \exp\left(t^2-y_t \exp\left(-t^2\right) \right) + z_t^2\exp\left(-t^2\right) \right)\,dt -z_t \,dW_t,\\  
   		   		 \quad y_T= \ln\left(\sin\left(W_T\right)+3\right) \exp\left(T^2\right).
   		   \end{cases}
  	\label{eq24}
\end{equation}
\label{ex2}
\end{exmp}
The analytic solution is 
\begin{equation}
   		  \begin{cases}
   		   		y_t = \ln\left(\sin\left(W_t\right)+3\right) \exp\left(t^2\right),\\  
   		   		z_t = \exp\left(t^2\right)\,\frac{\cos\left(W_t\right)}{\sin\left(W_t\right)+3}.
   		   \end{cases}
  	\label{eq25}
\end{equation}
The exact solution with $T=1$ is $\left(y_0,z_0\right)=\left(\ln\left(3\right),\frac{1}{3}\right)$. The naive results using $256$ threads per block with $K = K_y = K_z$, $t_0=0$, $T = 1$, $x \in [-16, 16]$, $L = 32$ and $p = 30$ are presented in Table~\ref{tab5} and plotted in Figure~\ref{fig2}.
\begin{table}[h!]
\centering
\caption{Naive results for Example~\ref{ex2}.}
\label{tab5}  
  \begin{tabular}{| c | c | c | c | c | c | c |c |}
    \hline
    K & N & M & $|y_{0,0}-y_0^0|$ & $|z_{0,0}-z_0^0|$ & $t_{CPU}$ & $t_{GPU}$ & speedup\\ \hline
    1  & 128 & 364 & 7.85E-04 & 3.52E-03 & 0.32 & 1.17 & 0.27 \\ \hline
    1  & 256 & 512 & 3.77E-04 & 1.76E-03 & 0.88 & 2.13 & 0.41 \\ \hline
    1  & 512 & 726 & 1.85E-04 & 8.78E-04 & 2.52 & 4.04 & 0.62 \\ \hline
    1  & 1024 & 1024 & 9.15E-05 & 4.39E-04 & 6.98 & 7.80 & 0.89 \\ \hline
    2  & 128 & 1218 & 1.85E-04 & 8.37E-04 & 1.52 & 1.24 & 1.23 \\ \hline
    2  & 256 & 2048 & 9.13E-05 & 4.29E-04 & 5.13 & 2.51 & 2.04 \\ \hline
    2  & 512 & 3446 & 4.54E-05 & 2.17E-04 & 17.47 & 5.11 & 3.42 \\ \hline
    2  & 1024 & 5794 & 2.26E-05 & 1.09E-04 & 58.93 & 10.65 & 5.53 \\ \hline
    3  & 128 & 4096 & 1.92E-07 & 8.34E-07 & 6.61 & 1.53 & 4.31 \\ \hline
    3  & 256 & 8192 & 2.41E-08 & 1.06E-07 & 26.81 & 2.97 & 9.03 \\ \hline
    3  & 512 & 16384 & 3.02E-09 & 1.33E-08 & 108.92 & 6.62 & 16.46 \\ \hline
    3  & 1024 & 32768 & 3.77E-10 & 1.67E-09 & 435.23 & 18.35 & 23.71 \\ \hline
    4  & 128 & 4096 & 1.10E-07 & 4.86E-07 & 8.06 & 1.53 & 5.28 \\ \hline 
    4  & 256 & 8192 & 1.42E-08 & 6.28E-08 & 32.82 & 3.02 & 10.87 \\ \hline
    4  & 512 & 16384 & 1.80E-09 & 7.99E-09 & 133.26 & 6.47 & 20.61 \\ \hline
    4  & 1024 & 32768 & 2.27E-10 & 1.01E-09 & 538.13 & 19.33 & 27.84 \\ \hline
    5  & 128 & 4096 & 1.20E-07 & 5.40E-07 & 9.48 & 1.54 & 6.14 \\ \hline
    5  & 256 & 8192 & 1.58E-08 & 7.04E-08 & 38.68 & 2.97 & 13.05 \\ \hline
    5  & 512 & 16384 & 2.02E-09 & 8.99E-09 & 156.63 & 6.67 & 23.48 \\ \hline
    5  & 1024 & 32768 & 2.55E-10 & 1.14E-09 & 635.01 & 19.55 & 32.48 \\ \hline
    6  & 128 & 4096 & 1.11E-07 & 5.08E-07 & 10.91 & 1.54 & 7.07 \\ \hline
    6  & 256 & 8192 & 1.49E-08 & 6.71E-08 & 44.77 & 3.09 & 14.48 \\ \hline
    6  & 512 & 16384 & 1.93E-09 & 8.63E-09 & 182.74 & 7.15 & 25.57 \\ \hline
    6  & 1024 & 32768 & 2.45E-10 & 1.09E-09 & 735.15 & 20.97 & 35.05 \\
    \hline
  \end{tabular}
\end{table}
We can observe that the accuracy for this example is smaller than the previous one. This is due to the convergence order of maximum 3, since the driver function depends on the $z$ process. Furthermore, we get higher speedup compared with previous example due to the more complicated driver function (i.e.\ more data are accessed, more special functional unit is used etc.). The naive speedup is $35\times$.

Furthermore, we optimized the kernels created for the this Example. We used the iterative optimization process described in Subsection~\ref{subsec42} for the case with $N=1024$ and $K_y=K_z=6$.   

\begin{table}
\centering
\caption{Results of iterative optimization process for Example~\ref{ex2}.}
\begin{subtable}[h!]{1\textwidth}
\centering
\caption{Performance of the main naive kernels.}
\begin{tabular}{| c | c | c|}
\hline 
Time(\%) & Time(s) & Kernel name\\ \hline
48.35 & 8.04 & nrm2\_kernel\\ \hline
14.94 & 2.48 & sp\_inter\_non\_grid\_d\_no\_for\\ \hline
13.70 & 2.28 & calc\_f\_and\_c\_exp\_d\\ \hline
6.17 & 1.03 & csrMv\_kernel\\ \hline
3.60 & 0.60 & calc\_y\\ \hline
3.53 & 0.89 & dot\_kernel\\ \hline
1.98 & 0.33 & reduce\_1Block\_kernel\\ \hline
1.56 & 0.26 & axpby\_kernel\_val\\ \hline
1.34 & 0.22 & calc\_c\_exp\_d\\ 
\hline
\end{tabular}
\label{tab6a}
\end{subtable}
\newline
\vspace*{1cm}
\newline
\begin{subtable}[h!]{1\textwidth}
\centering
\caption{Performance after first iteration of optimization process.}
\begin{tabular}{| c | c | c |}
\hline 
Time(\%) & Time(s) & Kernel name\\ \hline
27.88 & 2.49 & sp\_inter\_non\_grid\_d\_no\_for\\ \hline
25.53 & 2.28 & calc\_f\_and\_c\_exp\_d\\ \hline
11.35 & 1.01 & csrMv\_kernel\\ \hline
9.64 & 0.86 & dot\_kernel\\ \hline
6.74 & 0.60 & calc\_y\\ \hline
5.22 & 0.47 & reduce\_1Block\_kernel\\ \hline
2.65 & 0.24 & axpby\_kernel\_val\\ \hline
2.50 & 0.22 & calc\_c\_exp\_d\\ \hline
1.76 & 0.16 & step\_3\\ 
\hline
\end{tabular}
\label{tab6b}
\end{subtable}
\newline
\vspace*{1cm}
\newline
\begin{subtable}[h!]{1\textwidth}
\centering
\caption{Performance after second iteration of optimization process.}
\begin{tabular}{| c | c | c |}
\hline 
Time(\%) & Time(s) & Kernel name\\ \hline
22.23 & 1.46 & calc\_f\_and\_c\_exp\_d\\ \hline
17.67 & 1.16 & sp\_inter\_non\_grid\_d\_no\_for\\ \hline
15.58 & 1.02 & csrMv\_kernel\\ \hline
12.86 & 0.84 & dot\_kernel\\ \hline
9.05 & 0.60 & calc\_y\\ \hline
7.21 & 0.47 & reduce\_1Block\_kernel\\ \hline
3.41 & 0.22 & axpby\_kernel\_val\\ \hline
2.38 & 0.16 & step\_3\\ \hline
2.12 & 0.14 & copy\_d\\ 
\hline
\end{tabular}
\label{tab6c}
\end{subtable}
\label{tab6}
\end{table}

In the first iteration, we gathered the application information using $nvprof$. The results are presented in Table~\ref{tab6a}. The application hotspot is $nrm2\_kernel$ kernel, which calculates the second norm in the $BiCGSTAB$ algorithm. This is already optimized. Therefore, to overcome this bottleneck, we used the dot kernel $dot\_kernel$. The computation time is reduced from $8.04$ s to $0.86$ s. The new speedup after the first iteration is $57\times$.

In the second iteration, the new bottleneck for the application is the kernel that calculates the non-grid values for process $y$ and $z$ ($sp\_inter\_non\_grid\_d\_no\_for$) after each time layer backward. The performance of the kernel is limited by the latency of arithmetic and memory operations. Therefore, we considered loop interchanging and loop unrolling techniques. This reduced the computation time of the corresponding kernel and other kernels related with it, as shown in Table~\ref{tab6b}. We reduced the computation time from $2.48$ s to $1.16$ s for $sp\_inter\_non\_grid\_d\_no\_for$. By default, we have reduced the computation time from $2.28$ s to $1.46$ s for $calc\_f\_and\_c\_exp$ (the kernel in the third point of Subsection~\ref{subsec33}) because we needed to change the way how the non-grid points are stored and accessed and also reduction for $calc\_c\_exp\_d$ (calculates the conditional expectation) from $0.22$ s to $0.04$ s. The new speedup is $69\times$. It can be observed from Table~\ref{tab6c} that again the application bottleneck is the same kernel. Therefore, it is not worth optimizing the application furthermore. Finally, we decreased the block dimension from $256$ threads to $128$ in order to increase parallelism. The final speedup is $70\times$.  

In the following we consider an option pricing example, the Black-Scholes model. Consider a security market that contains one bond with price $p_t$ and one stock with price $S_t$. Therefore, their dynamics are described by:
\begin{equation}
   		  \begin{cases}
   		   		dp_t = r_t p_t \,dt,\quad t\ge0, \\  
   		   		 p_0 = p,
   		   \end{cases}
  	\label{eq26}
\end{equation}

\begin{equation}
   		  \begin{cases}
   		   		dS_t =  \mu_t S_t dt + \sigma_t S_t \,dW_t, \quad t \ge0,\\  
   		   		 \, \, S_0 = x,
   		   \end{cases}
  	\label{eq27}
\end{equation}
where $r_t$ denotes the interest rate of the bond, $p$ is its current value, 
$\mu_t$ is the expected return on the stock $S_t$, 
$\sigma_t$ is the volatility of the stock, 
$x$ is its current value and $W_t$ denotes the Brownian motion.

Suppose that an agent sells the option at price $y_t$ and then invests it in the market. Denote his wealth on each time by $y_t$. 
Assume that at each time the agent invests a portion of his wealth in an amount given by $\pi_t$ into the stock, and the rest $(y_t-\pi_t)$ into the bond. 
Now the agent has a portfolio based on the stock and the bond. 
Considering a stock that pays a dividend $\delta\left(t,S_t\right)$,
the dynamics of the wealth process $y_t$ are described by
\begin{equation}
\begin{split}
   	dy_t &= \frac{\pi_t}{S_t} \,dS_t 
   	      + \frac{y_t-\pi_t}{p_t} \, dp_t 
   	      + \pi_t\delta\left(t, S_t\right)\,dt  \\  
     	 &= \frac{\pi_t}{S_t}\left(\mu_t S_t\,dt + \sigma_t S_t\,dW_t\right) 
   	  + \frac{y_t-\pi_t}{p_t}\left( r_t p_t\,dt\right) + \pi_t\delta\left(t,S_t\right)\,dt  \\ 
   	  &= \left(r_t y_t + \pi_t \left(\mu_t - r_t + \delta\left(t, S_t\right)\right)\right)dt + \pi_t \sigma_t \,dW_t. 
\end{split}
  	\label{eq28}
\end{equation}
Let $z_t = \pi_t \sigma_t$, then
\begin{equation}
   	-dy_t = - \left( r_t y_t + \left(  \mu_t - r_t + \delta\left(t, S_t\right) \right)\frac{z_t}{\sigma_t} \right)\,dt + z_t \,dW_t.
  	\label{eq29}
\end{equation}
For a call option, one needs to solve a FBSDE, where the forward part is given from the SDE modelling of the stock price dynamics.
\begin{exmp}
Consider the Black-Scholes FBSDE
\begin{equation}
   		  \begin{cases}
   		   		\,\,\,\, dS_t =  \mu_t S_t\,dt + \sigma_t S_t \,dW_t, \quad S_0 = x,\\  
   		   	   -dy_t = - \left( r_t y_t + \left(  \mu_t - r_t + \delta\left(t, S_t\right) \right)\frac{z_t}{\sigma_t}\right)\,dt + z_t \,dW_t,\\  
   		   		\quad y_T = \left(S_T-K\right)^+.
   		   \end{cases}
  	\label{eq30}
\end{equation}
\label{ex3}
\end{exmp}
For constant parameters (i.e.\ $r_t = r$, $\mu_t = \mu$, $\sigma_t=\sigma$, $\delta_t = \delta$), the analytic solution is
\begin{equation}
   		  \begin{cases}
   		   		\quad y_t =  V\left(t,S_t\right)=S_t \exp\left(-\delta\left(T-t\right)\right) N\left(d_1\right)-K \exp\left(-r\left(T-t\right)\right) N\left(d_2\right),\\  
   		   	   \quad z_t = S_t \frac{\partial V}{\partial S}\sigma 
   		   	        = S_t \exp\left(-\delta \left(T-t\right)\right) N\left(d_1\right)\sigma, \\
   		   	   d_{1/2} = \frac{\ln\left(\frac{S_t}{K}\right) + \left( r \pm \frac{\sigma^2}{2} \right) \left(T-t\right)}{\sigma \sqrt{T-t}},
   		   \end{cases}
  	\label{eq31}
\end{equation}
where $N\left(\cdot\right)$ is the cumulative standard normal distribution function. 
In this example, we consider $T=0.33$, $K=S_0=100$, $r=0.03$, $\mu=0.05$, $\delta=0.04$ and $\sigma=0.2$ (taken from \cite{Zhao2010}) with the solution $\left(y_0,z_0\right)  \doteq \left(4.3671,10.0950\right)$.

Note that the terminal condition has a non-smooth problem for the $z$ process. 
Therefore, for discrete points near the strike price (also called at the money region), the initial value for the $z$ process will cause large errors on the next time layers. To overcome this non-smoothness problem, we considered smoothing the initial conditions, cf.\ the approach of Kreiss \cite{kreiss1970smoothing}. For the forward part of \eqref{eq31}, we have the analytic solution 
\begin{equation}
   	S_t = S_0\, \exp\Bigl(\bigl(\mu-\frac{\sigma^2}{2} \bigr)t + \sigma \,W_t\Bigr).  
  	\label{eq32}
\end{equation}
Discretizing \eqref{eq32}, the exponential term will lead to a non-uniform grid. 
Therefore, instead of working in the stock price domain, we work in the log stock price domain. If we denote $X_t = \ln\left(S_t\right)$, then the analytic solution of $X_t$ reads
\begin{equation}
   	X_t = X_0 + \bigl(\mu-\frac{\sigma^2}{2} \bigr)t + \sigma \,W_t.  
  	\label{eq33}
\end{equation}
The backward part is the same as in \eqref{eq30}. In Table~\ref{tab7} we show the importance of using the log stock price.

\begin{table}[h!]
\centering
\caption{Preliminary results for $N = 256, K_y = K_z = 3$ for Black-Scholes Example.}
\label{tab7}      
\begin{tabular}{| c | c | c | c |}
\hline 
$M$ & $t_{CPU}^{non-optimal}$ & $t_{CPU}^{optimal}$ & $speed up$  \\ \hline
24826 & 18531.23 & 47.54 & 389.80 \\
\hline
\end{tabular}
\end{table}

\begin{figure}[h!]
	\centering
	\begin{subfigure}[h!]{0.32\linewidth}
		\includegraphics[width=\linewidth]{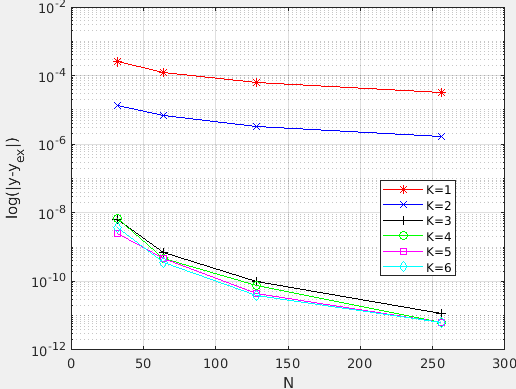}
		\caption{Plot of $y$ error as a function of time layers N for $K = 1,\cdots,6$.}
		\label{fig3a}
	\end{subfigure}
	\begin{subfigure}[h!]{0.32\textwidth}
		\includegraphics[width=\textwidth]{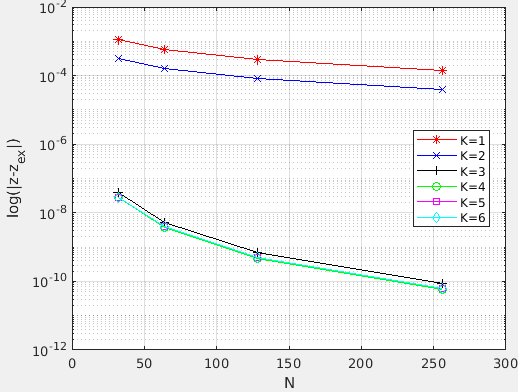}
		\caption{Plot of $z$ error as a function of time layers N for $K = 1,\cdots,6$.}
		\label{fig3b}
	\end{subfigure}
	\begin{subfigure}[h!]{0.32\textwidth}
		\includegraphics[width = \textwidth]{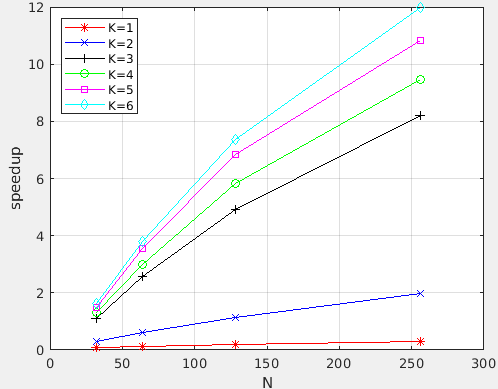}
		\caption{Plot of speedup as a function of time layers N for $K= 1,\cdots,6$.}
		\label{fig3c}
	\end{subfigure}
	\caption{Plots of naive results for Example~\ref{ex3}.}
	\label{fig3}
\end{figure}

 The naive results using $256$ threads per block with $K = K_y = K_z$, $t_0=0$, $T = 0.33$, $x \in [-16, 16]$, $L = 32$ and $p = 30$ are presented in Table~\ref{tab8} and plotted in Figure~\ref{fig3}.
\begin{table}[h!]
\centering
\caption{Naive results for Black-Scholes Example.}
\label{tab8}  
  \begin{tabular}{| c | c | c | c | c | c | c |c |}
    \hline
    K & N & M & $|y_{0,0}-y_0^0|$ & $|z_{0,0}-z_0^0|$ & $t_{CPU}$ & $t_{GPU}$ & speedup\\ \hline
    1  & 32 & 316 & 2.55E-04 & 1.11E-03 & 0.04 & 0.60 & 0.07 \\ \hline
    1  & 64 & 446 & 1.24E-04 & 5.70E-04 & 0.12 & 1.03 & 0.11 \\ \hline
    1  & 128 & 632 & 6.21E-05 & 2.89E-04 & 0.33 & 1.79 & 0.18 \\ \hline
    1  & 256 & 892 & 3.12E-05 & 1.45E-04 & 0.93 & 3.38 & 0.28 \\ \hline
    2  & 32 & 990 & 1.34E-05 & 3.12E-04 & 0.18 & 0.61 & 0.29 \\ \hline
    2  & 64 & 1664 & 6.88E-06 & 1.59E-04 & 0.64 & 1.04 & 0.61 \\ \hline  
    2  & 128 & 2798 & 3.38E-06 & 8.04E-05 & 2.16 & 1.92 & 1.13 \\ \hline
    2  & 256 & 4704 & 1.69E-06 & 4.04E-05 & 7.34 & 3.73 & 1.97 \\ \hline
    3  & 32 & 3104 & 6.45E-09 & 3.98E-08 & 0.70 & 0.63 & 1.11 \\ \hline
    3  & 64 & 6208 & 6.88E-10 & 5.35E-09 & 2.93 & 1.14 & 2.58 \\ \hline
    3  & 128 & 12414 & 9.72E-11 & 6.85E-10 & 11.81 & 2.39 & 4.93 \\ \hline
    3  & 256 & 24826 & 1.15E-11 & 8.50E-11 & 47.54 & 5.79 & 8.21 \\ \hline
    4  & 32 & 3104 & 6.86E-09 & 2.73E-08 & 0.85 & 0.64 & 1.32 \\ \hline
    4  & 64 & 6208 & 4.78E-10 & 3.81E-09 & 3.47 & 1.15 & 3.00 \\ \hline
    4  & 128 & 12414 & 7.55E-11 & 4.71E-10 & 14.26 & 2.45 & 5.82 \\ \hline
    4  & 256 & 24826 & 6.36E-12 & 5.93E-11 & 57.54 & 6.07 & 9.48 \\ \hline
    5  & 32 & 3104 & 2.55E-09 & 2.85E-08 & 0.94 & 0.64 & 1.48 \\ \hline
    5  & 64 & 6208 & 4.73E-10 & 4.05E-09 & 4.04 & 1.14 & 3.56 \\ \hline
    5  & 128 & 12414 & 4.40E-11 & 5.04E-10 & 16.30 & 2.39 & 6.83 \\ \hline
    5  & 256 & 24826 & 6.22E-12 & 6.41E-11 & 67.33 & 6.22 & 10.83 \\ \hline
    6  & 32 & 3104 & 3.77E-09 & 2.71E-08 & 1.06 & 0.65 & 1.64 \\ \hline
    6  & 64 & 6208 & 3.56E-10 & 3.90E-09 & 4.50 & 1.18 & 3.80 \\ \hline
    6  & 128 & 12414 & 3.82E-11 & 4.89E-10 & 18.69 & 2.54 & 7.35 \\ \hline
    6  & 256 & 24826 & 6.16E-12 & 6.24E-11 & 77.53 & 6.47 & 11.99 \\
    \hline
  \end{tabular}
\end{table}
The highest accuracy is achieved when considering a $6$-step scheme, and having also the highest speedup of $12\times$. 

We optimized the kernels created for the Black-Scholes BSDE for $N=256$ and $K_y=K_z=6$. The optimization iteration process is the same as in Example~\ref{ex2}. The final speedup is $31\times$. Note that this speedup is for $256$ time layers. In Example~\ref{ex2}, we optimized for $1024$ time layers. If we consider the same time layers for Black-Scholes Example, we get strange results (errors start to increase tremendously), due to the non-smooth problem of $z$ process. 

\begin{exmp}
Consider the $2$-dimensional BSDE\footnote{Taken from~\cite{Zhao2006}.}
\begin{equation}
   		  \begin{cases}
   		   		-dy_t =  \left( y_t - z_tA\right)\,dt - z_t\,dW_t,\\ 
   		   		 \quad y_T = \sin\left(MW_T+T\right),
   		   \end{cases}
  	\label{eq34}
\end{equation}
\label{ex4}
\end{exmp}
where $W_t = \left(W_t^1,W_t^2\right)^\top$, $z_t = \left(z_t^1,z_t^2\right)$, $A = \left(\frac{1}{2},\frac{1}{2}\right)^\top$ and $M = \left(1, 1\right)$.

\begin{figure}[h!]
	\centering
	\begin{subfigure}[h!]{0.32\linewidth}
		\includegraphics[width=\linewidth]{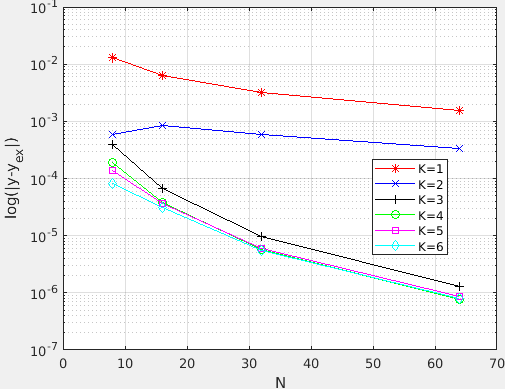}
		\caption{Plot of $y$ error as a function of time layers N for $K = 1,\cdots,6$.}
		\label{fig4a}
	\end{subfigure}
	\begin{subfigure}[h!]{0.32\textwidth}
		\includegraphics[width=\textwidth]{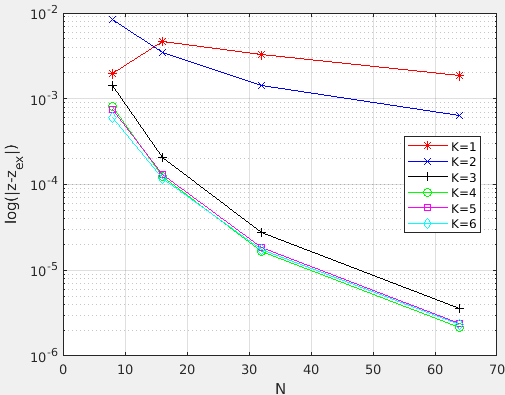}
		\caption{Plot of $z$ error as a function of time layers N for $K = 1,\cdots,6$.}
		\label{fig4b}
	\end{subfigure}
	\begin{subfigure}[h!]{0.32\textwidth}
		\includegraphics[width = \textwidth]{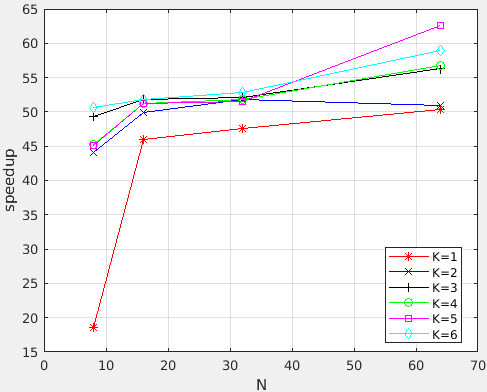}
		\caption{Plot of speedup as a function of time layers N for $K= 1,\cdots,6$.}
		\label{fig4c}
	\end{subfigure}
	\caption{Plots of naive results for Example~\ref{ex4}.}
	\label{fig4}
\end{figure}

The analytic solution is
\begin{equation}
   		  \begin{cases}
   		   	     y_t =  \sin\left(MW_t+t\right), \\ 
   		   		 z_t = \left(\cos\left(MW_t+t\right),\cos\left(MW_t+t\right)\right).
   		   \end{cases}
  	\label{eq35}
\end{equation}
The exact solution with $T=1$ is $\left(y_0,\left(z_0^1,z_0^2\right)\right)=\left(0,\left(1,1\right)\right)$. The naive results using $256$ threads per block with $K = K_y = K_z$, $t_0=0$, $T = 1$, $x \in [-8, 8]$, $L = 8$ and $p = 30$ are presented in Table~\ref{tab9} and plotted in Figure~\ref{fig4}.
\begin{table}[h!]
\centering
\caption{Naive results for Example~\ref{ex4}.}
\label{tab9}  
  \begin{tabular}{| c | c | c | c | c | c | c |c |c |c |}
    \hline
    K & N & M & $|y_{0,0}-y_0^0|$ & $|z_{0,0}-z_0^0|$ & $t_{CPU}$ & $t_{GPU}$ & speedup & Used GPU Memory (GB)\\ \hline
    1 & 8 & 46 & 1.32E-02 & 1.95E-03 & 0.14 & 0.01 & 18.60 & 0.30 \\ \hline
    1 & 16 & 64 & 6.46E-03 & 4.64E-03 & 0.53 & 0.01 & 45.96 & 0.30 \\ \hline
    1 & 32 & 92 & 3.18E-03 & 3.24E-03 & 2.12 & 0.04 & 47.55 & 0.30 \\ \hline
    1 & 64 & 128 & 1.57E-03 & 1.85E-03 & 8.73 & 0.17 & 50.33 & 0.30 \\ \hline
    2 & 8 & 78 & 5.90E-04 & 8.49E-03 & 0.62 & 0.01 & 44.08 & 0.30 \\ \hline
    2 & 16 & 128 & 8.31E-04 & 3.46E-03 & 3.56 & 0.07 & 49.93 & 0.31 \\ \hline
    2 & 32 & 216 & 5.84E-04 & 1.44E-03 & 21.96 & 0.42 & 51.75 & 0.34 \\ \hline
    2 & 64 & 364 & 3.39E-04 & 6.39E-04 & 131.09 & 2.57 & 50.95 & 0.42 \\ \hline
    3 & 8 & 128 & 3.94E-04 & 1.43E-03 & 2.12 & 0.04 & 49.40 & 0.32 \\ \hline
    3 & 16 & 256 & 6.75E-05 & 2.08E-04 & 21.05 & 0.41 & 51.83 & 0.38 \\ \hline
    3 & 32 & 512 & 9.72E-06 & 2.79E-05 & 187.79 & 3.61 & 52.04 & 0.65 \\ \hline
    3 & 64 & 1024 & 1.30E-06 & 3.60E-06 & 1719.55 & 30.54 & 56.30 & 1.66 \\ \hline
    4 & 8 & 128 & 1.90E-04 & 8.09E-04 & 2.41 & 0.05 & 45.19 & 0.33 \\ \hline
    4 & 16 & 256 & 3.78E-05 & 1.24E-04 & 25.99 & 0.51 & 51.18 & 0.40 \\ \hline
    4 & 32 & 512 & 5.69E-06 & 1.67E-05 & 239.73 & 4.63 & 51.80 & 0.75 \\ \hline
    4 & 64 & 1024 & 7.76E-07 & 2.17E-06 & 2279.06 & 40.10 & 56.83 & 2.09 \\ \hline
    5 & 8 & 128 & 1.39E-04 & 7.49E-04 & 2.39 & 0.05 & 45.06 & 0.35 \\ \hline
    5 & 16 & 256 & 3.65E-05 & 1.30E-04 & 30.16 & 0.59 & 51.27 & 0.43 \\ \hline
    5 & 32 & 512 & 6.00E-06 & 1.83E-05 & 292.10 & 5.67 & 51.54 & 0.86 \\ \hline
    5 & 64 & 1024 & 8.48E-07 & 2.42E-06 & 3049.33 & 48.75 & 62.55 & 2.52 \\ \hline
    6 & 8 & 128 & 8.13E-05 & 5.97E-04 & 2.21 & 0.04 & 50.61 & 0.34 \\ \hline
    6 & 16 & 256 & 3.07E-05 & 1.18E-04 & 32.93 & 0.64 & 51.76 & 0.46 \\ \hline
    6 & 32 & 512 & 5.49E-06 & 1.73E-05 & 341.15 & 6.46 & 52.81 & 0.97 \\ \hline
    6 & 64 & 1024 & 8.00E-07 & 2.32E-06 & 3394.13 & 57.56 & 58.96 & 2.94 \\ 
    \hline
  \end{tabular}
\end{table}
The highest speedup is ca. $59\times$, which requires a ca. $3$ GB of memory. Therefore, we did not consider more time layers as the maximum amount of memory for the GPU is $8$ GB and with $N=128$, we could not get the results for $K\ge3$. 
We optimized the case where $N=64$ and $K_y=K_z=6$.

\begin{table}[h!]
\centering
\caption{Performance of the main naive kernels for Example~\ref{ex4}.}
\label{tab10}
\begin{tabular}{| c | c | c|}
\hline 
Time(\%) & Time(s) & Kernel name\\ \hline
83.27 & 47.83 & main\_body\\ \hline
13.18 & 7.57 & swap\_coeff\\ \hline
3.23 & 1.86 & find\_coeff\\ \hline
0.16 & 0.91 & swap\_val\\ \hline
\end{tabular}
\end{table}
In the first iteration, we gathered the application information using $nvprof$. The results are presented in Table~\ref{tab10}. The application hotspot is $main\_body$ kernel, which calculates the non-grid values, conditional expectations and the values for process $y$ and $z$.  We merged these kernels compared with the $1$-dimensional case, due to memory constrain. Otherwise, we can't present results for $N=64$. The performance of the kernel is limited by the memory operations. Accessing of the bicubic spline coefficients is not optimal. However, we can't optimize this part. Moreover, we can't increase the block dimension because there are not enough resources. Therefore, the final speedup is ca. $59\times$. 

In the following we consider a $2$-dimensional option pricing problem, the zero strike European spread option. The BSDE is derived in the same way as in the Black-Scholes case. 

\begin{figure}[h!]
	\centering
	\begin{subfigure}[h!]{0.32\linewidth}
		\includegraphics[width=\linewidth]{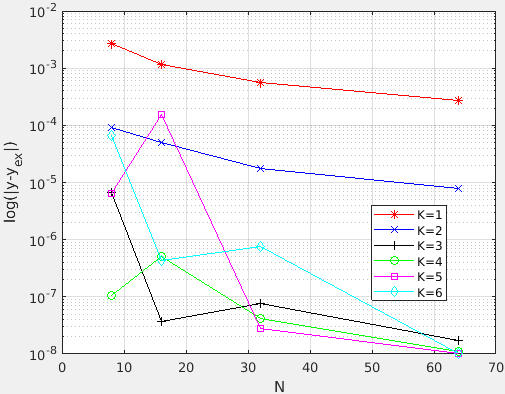}
		\caption{Plot of $y$ error as a function of time layers N for $K = 1,\cdots,6$.}
		\label{fig5a}
	\end{subfigure}
	\begin{subfigure}[h!]{0.32\textwidth}
		\includegraphics[width=\textwidth]{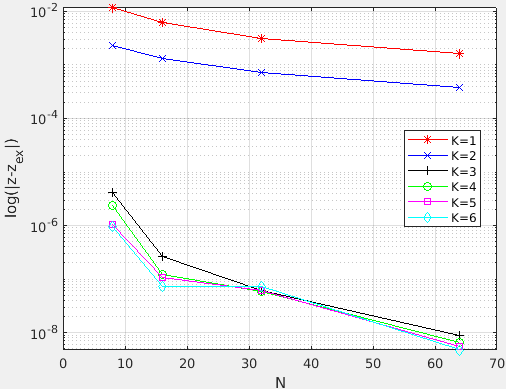}
		\caption{Plot of $z$ error as a function of time layers N for $K = 1,\cdots,6$.}
		\label{fig5b}
	\end{subfigure}
	\begin{subfigure}[h!]{0.32\textwidth}
		\includegraphics[width = \textwidth]{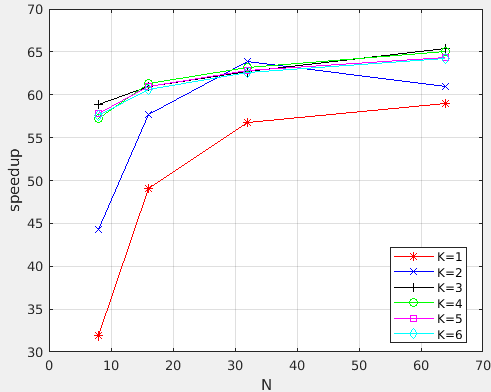}
		\caption{Plot of speedup as a function of time layers N for $K= 1,\cdots,6$.}
		\label{fig5c}
	\end{subfigure}
	\caption{Plots of naive results for Example~\ref{ex5}.}
	\label{fig5}
\end{figure}

\begin{exmp}
Consider the zero strike European spread option FBSDE
\begin{equation}
   		  \begin{cases}
   		   		\quad \, \, \, \, \, dS_t =  \mu S_t\,dt + \sigma S_t \,dW_t, \quad S_0 = x,\\ 
   		   		E\left[dW_t\right] = \rho d_t \\
   		   	   \quad \, \, -dy_t = - \left( r y_t + z_t A^{-1} M^{\top}\right)\,dt + z_t \,dW_t, \\  
   		   		\qquad \, \, y_T = \left(S_T^1-S_T^2\right)^+.
   		   \end{cases}
  	\label{eq36}
\end{equation}
\label{ex5}
\end{exmp}
where $S_t = \left(S_t^1,S_t^2\right)^\top$, $\mu = \left(\mu_1,\mu_2\right)$, $\sigma = \left(\sigma_1,\sigma_2\right)$, $W_t = \left(W_t^1,W_t^2\right)^\top$, $z_t = \left(z_t^1,z_t^2\right)$, $A =\begin{pmatrix}
\sigma_1 & 0 \\
\rho \sigma_2 & \sigma_2 \sqrt{1-\rho^2}
\end{pmatrix}$ and $M = \left(\mu_1 - r,\mu_2 - r\right)$.
The analytic solution is given by Margrabe's formula \cite{margrabe1978value}
\begin{equation}
   		  \begin{cases}
   		   		\quad y_t =  V\left(t,S_t\right)=S_t^1 N\left(d_1\right)-S_t^2 N\left(d_2\right),\\  
   		   	    \quad z_t = A^\top \nabla V S_t, \\
   		   	   d_{1/2} = \frac{\ln\left(\frac{S_t^1}{S_t^2}\right) \pm \frac{\tilde{\sigma}^2}{2} \left(T-t\right)}{\tilde{\sigma} \sqrt{T-t}},
   		   \end{cases}
  	\label{eq37}
\end{equation}
where $\tilde{\sigma} = \sqrt{\sigma_1^2 + \sigma_2^2 - 2\sigma_1\sigma_2\rho}$ and $\nabla V S_t= \left(\frac{\partial V}{\partial S_t^1}S_t^1,\frac{\partial V}{\partial S_t^2}S_t^2\right)^{\top}$. 
In this example, we consider $T=0.1$, $S_0^1=S_0^2=100$, $r=0.05$, $\mu_1=\mu_2=0.1$, $\sigma_1=0.25$, $\sigma_2=0.3$ and $\rho=0.0$ with the solution $\left(y_0,\left(z_0^1,z_0^2\right)\right) \doteq \left(15.48076,\left(14.3510,-12.6779\right)\right)$. The naive results are presented in Table~\ref{tab11} and plotted in Figure~\ref{fig5} for the same parameters as in Example~\ref{ex4}.
\begin{table}[h!]
\centering
\caption{Naive results for Example~\ref{ex5}.}
\label{tab11}  
  \begin{tabular}{| c | c | c | c | c | c | c |c |c |c |}
    \hline
    K & N & M & $|y_{0,0}-y_0^0|$ & $|z_{0,0}-z_0^0|$ & $t_{CPU}$ & $t_{GPU}$ & speedup & Used GPU Memory (GB)\\ \hline
    1 & 8 & 46 & 2.71E-03 & 1.21E-02 & 0.15 & 0.00 & 31.96 & 0.33\\ \hline
    1 & 16 & 64 & 1.16E-03 & 6.22E-03 & 0.61 & 0.01 & 49.03 & 0.34\\ \hline
    1 & 32 & 92 & 5.62E-04 & 3.18E-03 & 2.49 & 0.04 & 56.82 & 0.35\\ \hline
    1 & 64 & 128 & 2.67E-04 & 1.61E-03 & 10.29 & 0.17 & 58.93 & 0.34\\ \hline
    2 & 8 & 78 & 9.09E-05 & 2.31E-03 & 0.65 & 0.01 & 44.33 & 0.35\\ \hline
    2 & 16 & 128 & 5.06E-05 & 1.34E-03 & 4.08 & 0.07 & 57.72 & 0.35\\ \hline
    2 & 32 & 216 & 1.75E-05 & 7.22E-04 & 26.64 & 0.42 & 63.82 & 0.38\\ \hline
    2 & 64 & 364 & 8.00E-06 & 3.73E-04 & 153.21 & 2.51 & 60.96 & 0.46\\ \hline
    3 & 8 & 128 & 6.79E-06 & 4.23E-06 & 2.19 & 0.04 & 58.92 & 0.36\\ \hline
    3 & 16 & 256 & 3.73E-08 & 2.67E-07 & 22.90 & 0.38 & 61.02 & 0.42\\ \hline
    3 & 32 & 512 & 7.56E-08 & 6.22E-08 & 214.17 & 3.41 & 62.74 & 0.69\\ \hline
    3 & 64 & 1024 & 1.69E-08 & 9.01E-09 & 1911.74 & 29.26 & 65.34 & 1.70\\ \hline
    4 & 8 & 128 & 1.07E-07 & 2.40E-06 & 2.29 & 0.04 & 57.21 & 0.36\\ \hline
    4 & 16 & 256 & 5.16E-07 & 1.22E-07 & 28.10 & 0.46 & 61.37 & 0.45\\ \hline
    4 & 32 & 512 & 4.19E-08 & 5.90E-08 & 275.75 & 4.37 & 63.16 & 0.79\\ \hline
    4 & 64 & 1024 & 1.11E-08 & 6.46E-09 & 2509.67 & 38.55 & 65.10 & 2.13\\ \hline
    5 & 8 & 128 & 6.50E-06 & 1.06E-06 & 2.17 & 0.04 & 57.84 & 0.37\\ \hline
    5 & 16 & 256 & 1.53E-04 & 1.07E-07 & 32.14 & 0.53 & 60.92 & 0.47\\ \hline
    5 & 32 & 512 & 2.84E-08 & 6.07E-08 & 333.62 & 5.31 & 62.78 & 0.90\\ \hline
    5 & 64 & 1024 & 1.04E-08 & 5.54E-09 & 3083.75 & 47.94 & 64.32 & 2.56\\ \hline
    6 & 8 & 128 & 6.70E-05 & 9.72E-07 & 1.77 & 0.03 & 57.63 & 0.38\\ \hline
    6 & 16 & 256 & 4.27E-07 & 7.13E-08 & 35.05 & 0.58 & 60.59 & 0.50\\ \hline
    6 & 32 & 512 & 7.56E-07 & 7.14E-08 & 387.05 & 6.19 & 62.56 & 1.01\\ \hline
    6 & 64 & 1024 & 1.02E-08 & 4.73E-09 & 3666.74 & 57.06 & 64.26 & 2.99\\
    \hline
  \end{tabular}
\end{table}

\begin{table}[h!]
\centering
\caption{Performance of the main naive kernels for Example~\ref{ex5}.}
\label{tab12}
\begin{tabular}{| c | c | c|}
\hline 
Time(\%) & Time(s) & Kernel name\\ \hline
84.77 & 51.58 & main\_body\\ \hline
11.91 & 7.24 & swap\_coeff\\ \hline
3.00 & 1.83 & find\_coeff\\ \hline
0.14 & 0.88 & swap\_val\\ \hline
\end{tabular}
\end{table}

The highest speedup is ca. $64\times$, which requires ca. $3$ GB of memory. Note that the speedup is higher than in Example~\ref{ex4} since the driver function is more complex and more work is conducted from the GPU threads. Even here we couldn't optimize further as the main constrain is the GPU memory. In~\ref{tab12} we present the performance of the main naive kernels.
\newpage
\section{Conclusions and outlook}
\label{sec6}
In this work we parallelized the multistep method developed in \cite{teng2018multi} for solving BSDEs on GPU. Firstly, we presented an optimal operation to find the location of the interpolated values. This was essential for the reduction of the computational time. Our numerical results have shown that a high accuracy can be achieved with less computation times. For a further acceleration, we have investigated how to optimize the application after finding the performance bottlenecks and applying optimization techniques. The proposed GPU parallel algorithm make the multistep schemes in \cite{teng2018multi} and \cite{Zhao2010} be really more useful in practice. 
\clearpage
\bibliography{preprint}
\bibliographystyle{apalike}
\end{document}